\documentclass[prb,twocolumn,showpacs,preprintnumbers]{revtex4-1}
\usepackage{graphicx}
\usepackage{amsmath}
\usepackage{amssymb}
\usepackage{times}

\def\b{\beta}

\def\D{\Delta}
\def\d{\delta}
\def\e{\epsilon}

\def\S{\Sigma}
\def\s{\sigma}
\def\t{\tau}
\def\w{\omega}
\def\ua{\uparrow}
\def\da{\downarrow}

\def\Vec#1{\mathbf #1}
\begin{document}

\title{Doped high-$T_{\text{c}}$ cuprate superconductors \\
elucidated in the light of zeros and poles of electronic Green's function}

\author{Shiro Sakai,$^{1,2}$ Yukitoshi Motome,$^2$ and Masatoshi Imada$^2$}

\affiliation{$^1$Institute for Solid State Physics, Vienna University of Technology, 1040 Vienna, Austria\\
$^2$Department of Applied Physics, University of Tokyo, Hongo,Tokyo 113-8656, Japan}

\date{\today}

\begin{abstract}
We study electronic structure of hole- and electron-doped Mott insulators in the two-dimensional Hubbard model to reach a unified picture for the normal state
of cuprate high-$T_{\text{c}}$ superconductors. 
By using a cluster extension of the dynamical mean-field theory,
we demonstrate that structure of coexisting zeros and poles of
the single-particle Green's function holds the key to understand Mott physics in the underdoped region.
We show evidence for the emergence of non-Fermi-liquid phase caused by the topological quantum phase transition of Fermi surface by analyzing low-energy charge dynamics.
The spectra calculated in a wide range of energy and momentum
reproduce various anomalous properties observed in experiments for the high-$T_{\text{c}}$ cuprates. 
Our results reveal that the pseudogap in hole-doped cuprates
has a $d$-wave-like structure only below the Fermi level, while it retains non-$d$-wave structure with a fully opened gap above the Fermi energy even in the nodal direction due to a zero surface extending over the entire Brillouin zone.
In addition to the non-$d$-wave pseudogap,
the present comprehensive identifications of
the spectral asymmetry as to the Fermi energy, the Fermi arc, and the back-bending behavior of the dispersion, waterfall, and low-energy kink, in agreement with the experimental anomalies of the cuprates,  do not support that these originate from (the precursors of) symmetry breakings such as the preformed pairing and the $d$-density wave fluctuations, but support that
they are direct consequences of the proximity to the Mott insulator.
Several possible experiments are further proposed to prove or disprove our zero mechanism. 
\end{abstract}
\maketitle

%%%%%%%%%%%%%  Introduction  %%%%%%%%%%%%%
\section{INTRODUCTION}

Anomalous behaviors of high-$T_{\text{c}}$ cuprates observed in the normal metallic state above $T_{\text{c}}$ hold the key not only to understanding the mechanism of the superconductivity but also to a possible manifestation of an unexplored metallic phase distinguished from the Fermi liquid.\cite{ts99}
Extraordinary electronic structure is induced by a small density of carrier doping into the Mott insulator. 
Angle-resolved photoemission spectroscopies (ARPES) have in fact revealed detailed anomalies of the normal-state spectra,
such as momentum-dependent excitation gap (pseudogap), a truncated Fermi surface (Fermi arc), and kinks in the dispersion.\cite{dh03}

Toward the understanding of the anomalous metals, especially the pseudogap formation, many theoretical proposals have been made so far.\cite{yj03}
The proposals include a Cooper paring without phase coherence,\cite{ek95} 
and hidden orders or its fluctuations competing with the superconductivity, such as antiferromagnetism,\cite{ks90,p97,vt97} 
charge or stripe orderings,\cite{kb03} 
and $d$-density wave.\cite{v97,cl01}
Mechanisms attributing the origin of the pseudogap to a direct consequence of the proximity to the Mott insulator have also been 
proposed.\cite{rice05,sp03,sk06,p09,sm09}

Among the theoretical efforts, recent development of the dynamical mean-field theory (DMFT) \cite{gk96} and its cluster extensions \cite{ks01,mj05} has enabled studies on dynamics of microscopic models without any {\it ad hoc} approximation.
In particular, studies on the two-dimensional (2D) Hubbard model using the cluster DMFT (CDMFT) have offered many useful insights into the electronic structure of cuprates, by identifying the pseudogap, Fermi arc,\cite{mp02,st04,sm09,cc05,kk06,sk06,mj06,aa06,hk07,lt09,fc09} and high-energy kink \cite{mj07} in the calculated spectra.
The CDMFT is specifically suited and powerful for this problem because of its nonperturbative
framework, namely, it is based on neither weak nor strong coupling expansions.  
Furthermore, 
it takes account of short-range spatial correlations within a cluster explicitly.
These are big advantages in exploring momentum-resolved dynamics in the intermediate coupling region, which is relevant to physics of the cuprates. 
In particular, recent CDMFT studies on doped Mott insulators have revealed emergence of non-Fermi-liquid phases, characterized by unexpected coexistence of two singularities at the Fermi level, one characteristic to the weak-coupling and the other to the strong-coupling regions.\cite{sk06,sm09,sm09-2,lt09,wg09}

\begin{figure}[b]
\center{
\includegraphics[width=0.48\textwidth]{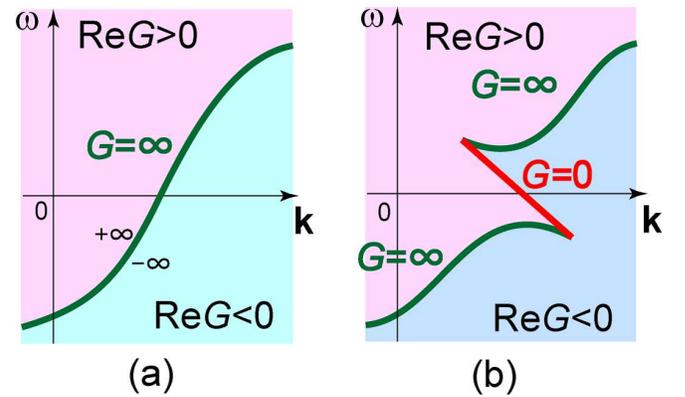}}
\caption{(color online). Schematics of the sign of Re$G$ for 
       (a) normal metals and (b) correlation-driven insulating states. 
       $\Vec{k}$ is the momentum and 
       $\w$ is the energy measured from the Fermi level.}
\label{fig:zero}
\end{figure}

The singularity characterizing weakly interacting metals, namely, 
Fermi liquids, is a pole of the single-particle Green's function $G$.  
Energy dependence of its locus in the momentum space 
determines band dispersions, particularly, 
the Fermi surface at the Fermi level. 
On the other hand, a crucial singularity in the strong coupling region is a zero of $G$. 
The situations are illustrated in Fig.~\ref{fig:zero}, which schematically shows how the real part of $G$ changes the sign in the energy-momentum space.
Because Re$G$ must be positive at high energy while negative at low energy, it has to change its sign at least once between these two regions.
The sign change also has to take place in the Brillouin zone
at the Fermi level, say, between the zone center and the zone boundary, 
if the dispersion far from the Fermi level ensures positive Re$G$ in some part of momenta and negative Re$G$ in the other part. 
In the normal metals, the sign change occurs at the poles of $G$ (i.e., the band dispersions), where Re$G$ goes to $-\infty$ on the lower side of the band and comes back from $+\infty$ on the upper side [Fig.~\ref{fig:zero}(a)].
On the other hand,
what happens when correlation effects induce a gap in the band at the Fermi level?
Because of the absence of poles inside the gap by definition, 
there is no way for Re$G$ to change its sign at the Fermi level except for getting through zeros [Fig.~\ref{fig:zero}(b)].\cite{d03,sp07} 
Re$G=0$ means the divergence of the self-energy, so that a zero of $G$ is a singular point of the self-energy.

The non-Fermi liquids discovered in CDMFT show a coexistence of these two characteristics --- poles and zeros of $G$ at the Fermi level.\cite{sk06,sm09}
In a lightly hole-doped region, the poles form a hole-pocket Fermi surface around the nodal direction [from $\Vec{k} = (0,0)$ to $(\pi,\pi)$], 
while the zeros form a surface enclosing $(\pi,\pi)$.
It has been suggested that the Fermi arc emerges because of such peculiar electronic structure: The pocket loses the spectral intensity on the side closer to the zero surface, leaving the other side as an arc.\cite{sk06,sm09,sm09-2}
Similar mechanisms were proposed, based on an assumed
functional form of Green's functions \cite{kr06,ks06} or 
a weakly-coupled chain model.\cite{et02,bg06}
The pseudogap is also characterized by the same zero surface crossing the Fermi level.
In the previous paper \cite{sm09} we clarified the structure of poles and zeros in the whole energy-momentum space for hole-doped and undoped Mott insulators.  We thus proposed a unified picture for understanding various puzzling features of Mott physics in the light of reconstructions of pole-zero structure under progressive doping from the Mott insulator to the Fermi liquid in the overdoped region.

The aim of the present paper is to show how and to what extent we can understand experimental findings within the `pole-zero mechanism'.
We implement CDMFT calculations at zero and finite temperatures,  
clarifying the relationship between zero-temperature pole-zero structure and the spectra observed in experiments at finite temperatures.
We find that a number of non-Fermi-liquid aspects, i.e., 
spectral asymmetry as to the Fermi level, back-bending or incoherent feature of the dispersion, high- and low-energy kinks, and Fermi arc or pockets in the hole- and electron-doped cuprates, 
are comprehensively understood within the pole-zero mechanism.
In addition to a coherent picture for various experiments, 
our numerical results predict a distinctive feature of the pseudogap:
The numerical data support a fully opened gap above the Fermi level even in the nodal direction, which is incompatible with scenarios on the basis of the $d$-wave gap in the zero-temperature limit.
Our result offers a mechanism distinct from the scenarios based on the preformed pairing or $d$-density-wave order.

%-------- organization
The paper is organized as follows.
In Sec.~\ref{sec:method} we introduce the 2D Hubbard model and present the central ideas of the CDMFT.
We also describe the essence of numerical solvers for the effective cluster problem that we used in the present study. 
The numerical results are presented and discussed in comparison with various experiments in Sec.~\ref{sec:result}.
From the low-energy pole-zero structure, 
we derive a simple interpretation of several unusual features in 
hole-doped cuprates (Sec.~\ref{ssec:nond}-\ref{ssec:back}).
We also find several anomalies in the band dispersion, which are 
consistent with ARPES data (Sec.~\ref{ssec:edc}-\ref{ssec:kink}).
In Sec.~\ref{ssec:arc} we discuss the results of the Fermi surface 
in comparison with the recent ARPES observation of the hole pocket.
The comparison is further made in Sec.~\ref{ssec:tp} 
by changing the next-nearest-neighbor hopping to reach more 
quantitative understanding of the experimental results. 
In Sec.~\ref{ssec:tp}, \ref{ssec:ele2}, and \ref{ssec:ele}, we extend our theory to electron-doped cuprates and find good agreements with ARPES data.
We summarize our results and make concluding remarks in Sec.~\ref{sec:summary}.

%%%%%%%%%%%%%  Model & Method  %%%%%%%%%%%%%
\section{MODEL AND METHOD} \label{sec:method}

As a simplest model for high-$T_{\text{c}}$ cuprates we take the Hubbard Hamiltonian,
\begin{align}
H= \sum_{\Vec{k}\s}\e(\Vec{k})c_{\Vec{k}\s}^\dagger c_{\Vec{k}\s}
-\mu \sum_{i\s}n_{i\s}+ U\sum_{i}n_{i\ua}n_{i\da},
\label{eq:hubbard}
\end{align}
on a square lattice.
Here $c_{\Vec{k}\s}$ $(c_{\Vec{k}\s}^\dagger)$ annihilates (creates) 
an electron of spin $\s$ with momentum $\Vec{k} = (k_x, k_y)$, 
$c_{i\s}$ $(c_{i\s}^\dagger)$ is its Fourier component at site $i$, and
$n_{i\s}\equiv c_{i\s}^\dagger c_{i\s}$.
$U$ represents the onsite Coulomb repulsion, $\mu$ the chemical
potential, and 
\begin{align}
\e(\Vec{k})\equiv -2t(\cos k_x + \cos k_y) -4t' \cos k_x\cos k_y,
\label{eq:disp}
\end{align}
where $t$ $(t')$ is the (next-)nearest-neighbor transfer integral. 

Based on a first principles calculation, 
the value of $t$ was estimated to be $\sim 0.4$eV for La$_2$CuO$_4$.\cite{hs90}
$-t'/t$ is considered to be $\sim 0.2$ and $\sim 0.4$ for La$_{2-x}$Sr$_x$CuO$_4$ and Bi$_2$Sr$_2$CaCu$_2$O$_{8+\d}$, respectively.
We adopt $U=8t$ or $12t$, which are realistic values for the cuprates and indeed reproduce the Mott insulating state for undoped case.

In the CDMFT \cite{ks01} we map the system (\ref{eq:hubbard}) onto a model consisting of an $N_{\text{c}}$-site cluster C and bath degrees of freedom B.
The bath is determined in a self-consistent way to provide an $N_{\text{c}}\times N_{\text{c}}$ dynamical mean-field matrix $\hat{g}_0(i\w_n)$ at temperature  $T(\equiv 1/\b)$, where $\w_n\equiv (2n+1)\pi T$. 

After the self-consistency loop converges, we calculate a quantity 
$Q^{\text{L}}$ defined on the original lattice from those on the cluster, $Q^{\text{C}}$.\cite{ks01}
This periodization procedure is based on the Fourier transformation truncated by the cluster size $N_{\text{c}}$,
\begin{align}
Q^{\text{L}}(\Vec{k})=\frac{1}{N_{\text{c}}}\sum_{ij\in \text{C}}
                     [Q^{\text{C}}]_{ij}e^{i\Vec{k}\cdot\Vec{r}_{ij}},
\label{eq:periodize}
\end{align}
where $\Vec{k}$ is defined on the entire Brillouin zone 
of the original lattice
and $\Vec{r}_{ij}$ is the real-space vector connecting two cluster sites $i$ and $j$.

The truncation by a small $N_{\text{c}}$ gives a good approximation to the thermodynamic limit of $N_{\text{c}} \to \infty$ if $Q^{\text{C}}$ is well localized within the cluster.
In reality it is difficult to find a quantity which is short ranged
in the entire parameter range of the interaction strength and the doping concentration.
Therefore we need to choose an appropriate quantity to periodize, according to situations.
For example, $Q=\S$ ($\S$: self-energy)
 is a good choice for weakly interacting normal metals,\cite{ks01}  
but it becomes highly nonlocal and long ranged in the strong coupling region (e.g., in the Mott insulator). 
This is due to an appearance of zeros of $G$, i.e., poles of $\S$ in the momentum space (see APPENDIX A for further detail).
On the other hand, in the Mott insulator $Q=G$ is more appropriate because 
it is nearly local in the strong coupling regime.\cite{kk06}

Another choice for the periodization is the cumulant 
\begin{equation}
M=[i\w_n+\mu-\S]^{-1}.
\end{equation}
It was pointed out that the cumulant periodization works well 
in a wide range of $U$ including the strong coupling, 
because it is similar in the functional form to the atomic Green function.\cite{sk06} 
Moreover, the periodization by the cumulant has an important feature, i.e., 
it can describe both poles and zeros of $G$ at the same time, while 
the periodization by using $\S$ ($G$) describes only poles (zeros). 
This opens up the intriguing possibility of exploring the coexistence of poles and zeros at the Fermi level, as substantiation of anomalous metals.  

In the present study, we adopt the cumulant periodization scheme, $Q=M$, to investigate {\it metals in the vicinity of the Mott insulator}. 
In fact, $M$ is highly local even in the doped metallic states.
In APPENDIX B we demonstrate this local nature by 
CDMFT calculations for $N_{\text{c}}=4\times4$ cluster; 
we find that $M^{\text{C}}$ is nearly localized already within the inner $2\times 2$ cluster for a parameter region relevant to the present study.
After obtaining the lattice cumulant $M^{\text{L}}$ through Eq.~(\ref{eq:periodize}), we calculate the self-energy $\S^{\text{L}}$, 
Green's function $G^{\text{L}}$, and spectral function $A^{\text{L}}$ on the original lattice with
\begin{align}
\S^{\text{L}}(\Vec{k},\w)\equiv 
\left[\w+\mu-{M^{\text{L}}}^{-1}(\Vec{k},\w)\right]^{-1},\nonumber\\
G^{\text{L}}(\Vec{k},\w)\equiv 
\left[\w+\mu-\e(\Vec{k})-\S^{\text{L}}(\Vec{k},\w)\right]^{-1},
\end{align}
and
\begin{align}
A^{\text{L}}(\Vec{k},\w)\equiv 
-\frac{1}{\pi}\text{Im}G^{\text{L}}(\Vec{k},\w).
\end{align}
In the following calculations, we employ an $N_{\text{c}}=2\times 2$ cluster, 
and concentrate on the paramagnetic metallic solution. 

We numerically solve the effective cluster problem by means of the exact diagonalization (ED) method at $T=0$ and 
the continuous-time quantum Monte Carlo (CTQMC) method at $T>0$, as we elaborate below.
We hereafter omit the superscript L in $\S^{\text{L}}$, $G^{\text{L}}$ and $A^{\text{L}}$. 

\subsection{Exact diagonalization method}\label{ssec:ed}

Although the pseudogap state in cuprates is experimentally detectable only above $T_{\text{c}}$, it is still significant to elucidate
the nature of the pseudogap state in the zero temperature limit, 
by assuming the paramagnetic metal for the doped Mott insulator. 
This is a circumstance similar to the normal metal, where the concept of the Fermi liquid justified only in the zero temperature limit in the strict sense has proven to be fruitful. 
For this purpose of clarifying the zero temperature limit, 
we employ the Lanczos ED method \cite{l50} for the cluster problem.

In this scheme we take a large but finite value of pseudo inverse temperature 
$\b'$ ($=100/t$ or $200/t$ throughout the paper), which practically represents the ground state with an energy resolution corresponding to $1/\b'$. 
We then represent the dynamical mean field $\hat{g}_0$ with a finite number $N_{\text{B}}$ of bath degrees of freedom, which constitute together with C sites the effective Hamiltonian to be diagonalized.
We take $N_{\text{B}}=8$ throughout the paper.

The optimization of B sites is done by minimizing the distance function defined by
\begin{align}
d\equiv \sum_{ij\in \text{C}}\sum_n 
     \left|\left[\hat{g}_0(i\w_n)\right]_{ij}
     -\left[\hat{g}_{0,N_{\text{B}}}(i\w_n)\right]_{ij}\right|^2 
     e^{-\w_n/t},
\label{eq:distance}
\end{align}
where $\hat{g}_{0,N_{\text{B}}}$ is the non-interacting Green's function for the effective Hamiltonian and we have introduced the exponential weight factor $e^{-\w_n/t}$ with $\w_n\equiv (2n+1)\pi/\b'$ 
to reproduce more precisely the important low-frequency part.
We have examined several other types of distance functions and confirmed that qualitative feature of the results obtained in this paper does not depend on the choice.

An advantage of the Lanczos method is that Green's function 
is obtained as a function of real frequency $\w$ by the continued-fraction expansion,
\begin{align} 
G(\w)&=\frac{\langle 0| c c^\dagger | 0\rangle}
   {\w+i\eta-a_0^>-\displaystyle{ \frac{{b_1^>}^2}{\w+i\eta-a_1^>
    - \displaystyle{\frac{{b_2^>}^2}{\w+i\eta-a_2^>-\cdots}}}}}\nonumber\\
  &+\frac{\langle 0| c^\dagger c | 0\rangle}
   {\w+i\eta-a_0^<-\displaystyle{\frac{{b_1^<}^2}{\w+i\eta-a_1^<
    -\displaystyle{\frac{{b_2^<}^2}{\w+i\eta-a_2^<-\cdots}}}}},
\label{eq:cf}
\end{align}
where $|0\rangle$ is the ground-state vector and the coefficients $a_i^{>,<}$ and $b_i^{>,<}$ are the elements of the tridiagonal matrix appearing in the Lanczos algorithm.\cite{gb87}
We take account of up to 2000th order in the expansion.
A small positive $\eta$ is introduced to satisfy the causality.
In principle, $\eta$ is taken to be infinitesimal, 
but in practice, it is useful to consider $\eta$ as a parameter, which serves as a resolution in energy or mimics an infinite-size effect not incorporated into the ED calculation.

In the following study, we pursue a further benefit of the parameter $\eta$: 
We use $\eta$ as a mimic of the source of incoherence in the electronic structure, 
such as thermal or impurity scattering. 
This is based on the observation that $\eta$ does not substantially change the location of poles and zeros of $G$ but changes only the sharpness of these singularities.\cite{sm09,sm09-2}
Indeed, as long as $\eta$ is sufficiently smaller than the typical energy scale of the system, the location of the poles and zeros is virtually determined only by $a_i^{>,<}$ and $b_i^{>,<}$ in Eq.~(\ref{eq:cf}), which are calculated in the self-consistency loop performed on the Matsubara-frequency axis and independent from $\eta$.
Thus, $\eta$ provides an opportunity to get insight into how the electronic structure at finite temperatures evolves from the pole-zero structure at zero temperature. 
The effect of $\eta$ is confirmed by a direct comparison 
with the finite-temperature results obtained by QMC introduced below. 
Note that this smearing technique by $\eta$ is very useful 
because it is difficult for QMC to obtain the precise electronic structure
at real frequencies because it requires an analytic continuation of the numerical data.
We use the smearing technique in Sec.~\ref{sec:result} to compare the CDMFT+ED results with ARPES ones.

\subsection{Quantum Monte Carlo method}\label{ssec:ctqmc}

In order to discuss thermal effects directly, we implement QMC 
calculations for the effective cluster problem.
Since the scheme takes into account infinite bath degrees of freedom, the results complement the limitation in the bath size in the CDMFT+ED results as well.
We adopt the algorithm based on a weak-coupling series expansion and auxiliary-field transformation.
The idea was first proposed by Rombouts {\it et al.},\cite{rh99} applied to DMFT by Sakai {\it et al.},\cite{sa06} and recently formulated in a sophisticated fashion by Gull {\it et al.}\cite{gw08}, which enabled a study at sufficiently low temperatures.

In the QMC, we first expand the many-body partition function in the interaction part of the Hamiltonian and apply the Hubbard-Stratonovich decoupling \cite{rh98,h83} to the interaction part.
This decomposes the many-body partition function into a sum of single-particle systems, which we collect.

While the algorithm is based on the series expansion up to infinite order,
it is feasible to obtain a numerically exact result because, 
after numerical convergence, all the orders in the expansion are virtually 
taken into account.\cite{rh99}

We implement the CTQMC sampling for the auxiliary fields, employing the updating algorithm proposed in Ref.~\onlinecite{gw08}.
We typically take $2\times10^5$ steps for each QMC simulation, and after convergence in the self-consistency loop, 
we average over 30 data starting from $g_0$'s which are different within a statistical error bar.

%%%%%%%%%%%%%  Result & Discussion  %%%%%%%%%%%%%

\section{RESULT AND DISCUSSION}\label{sec:result}

\subsection{Non-{\it d}-wave pseudogap}\label{ssec:nond}

\begin{figure}[t]
\center{
\includegraphics[width=0.48\textwidth]{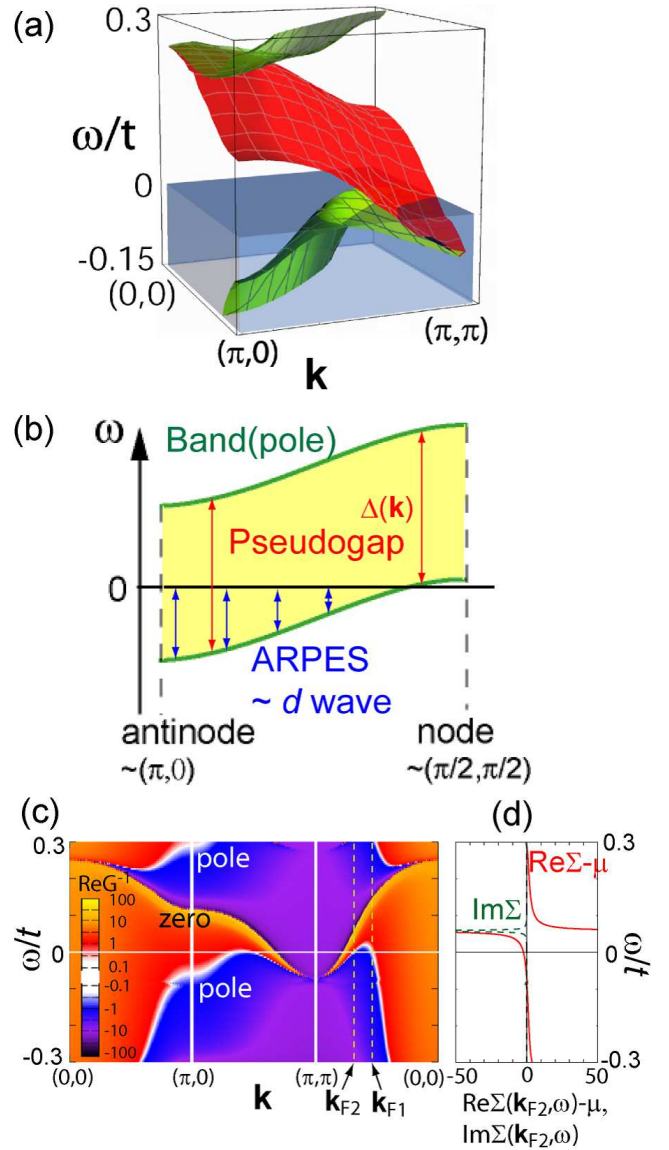}}
\caption{(color). 
       (a) Low-energy structure of poles and zeros of $G$ 
       for $t'=0$, $U=8t$, and $n=0.91$.
       Reproduction of Fig.~2(a) in Ref.~\onlinecite{sm09}.
       (b) A schematic picture for the relation between 
       the calculated pseudogap and that observed in ARPES.
       The left side represents the antinodal region around $(\pi,0)$
       while the right side the nodal region around 
       $(\frac{\pi}{2},\frac{\pi}{2})$.
       (c) Re$G^{-1}(\Vec{k},\w)$ plotted against energy and momentum 
       along the symmetric lines of $\Vec{k}$ in the Brillouin zone. 
       The parameters are the same as in (a).
       White curves between blue and red region represent poles of $G$ 
       while boundaries between black and yellow region represent zeros. 
       $\Vec{k}_{\text{F1}}$ and $\Vec{k}_{\text{F2}}$ represent 
       the Fermi wave numbers at the hole pocket along the nodal line. 
       (d) Energy dependence of the self-energy at the Fermi momentum
       $\Vec{k}_{\text{F2}}$ denoted in (c).
       }
\label{fig:fig2}
\end{figure}

First of all, we make a remark on the structure of pseudogap obtained in Fig.~\ref{fig:fig2}(a) in Ref.~\onlinecite{sm09} [reproduced in Fig.~\ref{fig:fig2}(a)], which depicts the low-energy pole and zero surfaces calculated 
by CDMFT+ED for 9\% hole doping to the Mott insulator at $U=8t$ and $t'=0$.
The pseudogap is formed by the zero surface (red) connecting two separated pole surfaces (green). 
Here we define the $\Vec{k}$-dependent amplitude $\D(\Vec{k})$ of the pseudogap as the energy difference between the upper and the lower poles at each $\Vec{k}$.
Then we see in Figs.~\ref{fig:fig2}(a) and (c) that 
the gap in the direct transition opens in the entire Brillouin zone,
 though it is somewhat larger in the antinodal region [$\D((\pi,0))=0.35t$] than in the nodal region 
[$\D(\Vec{k}_{\text{F1}}\equiv(0.55\pi,0.55\pi))=0.31t$]. 
This is one of the central results of this paper, and clearly different from previous scenarios of pseudogap which assume a $d$-wave gap above $T_{\text{c}}$, such as preformed pair,\cite{ek95} {\it d}-density-wave,\cite{v97,cl01} 
resonating valence-bond,\cite{rice05,kr06}
and nodal-liquid theory,\cite{bf98,ft01} since 
the direct transition gap closes 
in the nodal direction in these scenarios.

Nevertheless the electronic structure in Fig.~\ref{fig:fig2}(a), which was calculated without any assumption on the gap structure for the microscopic model, 
is consistent with the ARPES data, as described below.
To start with, it should be noted that ARPES observes only the spectra below the Fermi energy $E_{\text{F}}(\equiv 0)$ if $T$ is low, 
so that the gap amplitude is often estimated by symmetrizing the spectra below and above $E_{\text{F}}$.\cite{nd98}
This is a misleading, artificial procedure because it assumes a symmetric structure of the gap as to the Fermi energy and neglects the structure above $E_{\text{F}}$ if any. 
Suppose we have the result in Fig.~\ref{fig:fig2}(a) only below $E_{\text{F}}$ and symmetrize it as in the ARPES procedure to estimate the pseudogap, 
we end up with a $d$-wave like gap since 
the gap in the part
below $E_{\text{F}}$ is larger in the antinodal region while it is smaller or even zero in the nodal region
[this is more clearly seen in Fig.~\ref{fig:fig2}(c) where we plot Re$G^{-1}(\Vec{k},\w)$ against energy and momentum along symmetric lines].
The situation is schematically shown in Fig.~\ref{fig:fig2}(b). 
Our results suggest that the ``$d$-wave structure" is an artifact of the 
symmetrizing analysis, and in reality, the pseudogap has a non-$d$-wave (namely, full-gap) structure.
In APPENDIX C we confirm that the 
gap in the nodal direction persists for a larger cluster, by implementing an $N_\text{c}=8$ CDMFT+CTQMC calculation at a low temperature.

It is interesting to examine this interpretation experimentally. 
Our theory predicts that a 
 gap exists above $E_{\text{F}}$ even in the nodal direction 
if the paramagnetic metal persists down to zero temperature. 
This means that the symmetrization \cite{nd98} used in ARPES breaks down in the pseudogap state due to the spectral asymmetry as to the Fermi energy. 
This may be examined by spectroscopic or scattering probes to observe unoccupied electronic states above the Fermi level, such as the inverse photoemission spectroscopy (IPES). 
It is highly desired to reveal the pseudogap structure by combining both PES and IPES without any symmetrization procedure. 
Other possible experimental probes suited for this purpose may be electron energy loss spectroscopy, resonant inelastic X-ray scattering, and time-resolved photoemission.
In addition, another possible study is ARPES on electron-doped cuprates.
When we interpret the result with the electron-hole transformation, it provides information on the spectra of unoccupied states in a hole-doped system. 
We will discuss this in Sec.~\ref{ssec:ele2} and \ref{ssec:ele}, 
comparing the results with existing ARPES data on electron-doped cuprates.

\subsection{Spectral symmetry and asymmetry around the Fermi energy}\label{ssec:eh}

Although a thorough comparison of the calculated pole-zero structure with
experiments is not possible at present because of the lack of the experimental
spectra above $E_{\text{F}}$,
it is still significant to make comparison with available experimental data 
which partly elucidated them.
The scanning tunnelling microscopy (STM) \cite{hl04} and the ARPES \cite{yr08} 
reported such data, where they found an electron-hole asymmetry of the spectra around $E_{\text{F}}$.
The asymmetry provides a clue to the mechanism of the pseudogap, especially to the relation between the preformed pairing and the pseudogap, because the pairing will lead to a symmetric spectrum as established in the Bardeen-Cooper-Schrieffer (BCS) theory.\cite{bc57}

\begin{figure}[t]
\center{
\includegraphics[width=0.48\textwidth]{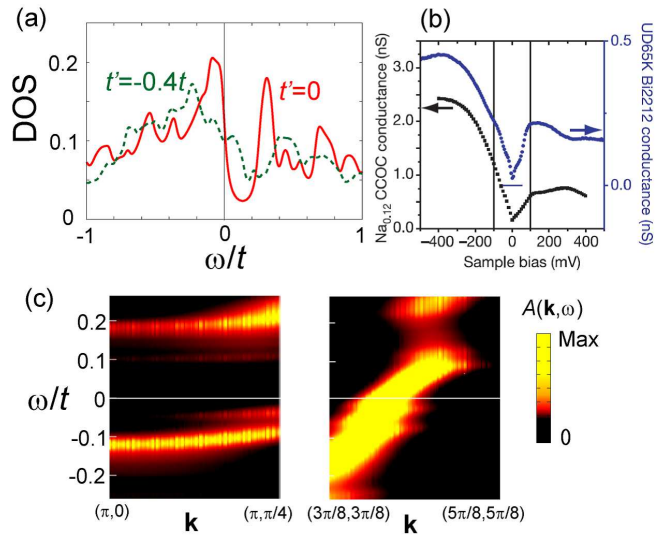}}
\caption{(color online). (a) DOS for $U=8t$, $t'=0$, and $n=0.94$ (solid curve)           
         and for $U=8t$, $t'=-0.4t$, and $n=0.95$ (dashed curve).
       (b) Spatially averaged tunnelling conductance spectrum of lightly 
       hole-doped cuprates, Na$_{0.12}$Ca$_{1.88}$CuO$_2$Cl$_2$ (left scale)
       and Bi$_2$Sr$_2$CaCu$_2$O$_{8+\d}$ (right scale). 
       Reproduction of Fig.~1(c) in Ref.~\onlinecite{hl04}.
       (c) Spectral function calculated with the CDMFT+ED
       around the antinode $\left[(\pi,0)\text{-}(\pi,\frac{\pi}{4})\right]$ 
       and the node 
       $\left[(\frac{3\pi}{8},\frac{3\pi}{8})
             \text{-}(\frac{5\pi}{8},\frac{5\pi}{8})\right]$ 
       for $t'=-0.4t$, $U=8t$, and $n=0.95$.
       The numerical data are broadened with $\eta=0.03t$.
       }
\label{fig:asym}
\end{figure}

Figure \ref{fig:asym}(a) shows the low-energy part of the density of states (DOS) for $U=8t$, $t'=0$, and $n=0.94$ and for $U=8t$, $t'=-0.4t$, and $n=0.95$.
The data show that the weight of the low-energy occupied states is significantly larger than that of unoccupied ones,
in accord with other CDMFT studies.\cite{sk06,kk06}
The asymmetry is consistent with the spatially averaged electron-tunnelling spectra [Fig.~\ref{fig:asym}(b), reproduced from Fig.~1(c) in Ref.~\onlinecite{hl04}] measured by the STM for lightly hole-doped cuprates, Na$_{0.12}$Ca$_{1.88}$CuO$_2$Cl$_2$ and Bi$_2$Sr$_2$CaCu$_2$O$_{8+\d}$, where the observed probability of electron extraction is considerably greater than that of injection.
Moreover, the numerical data show a peak around $\w=0.3t(0.4t)$ for $t'=0(-0.4t)$, which roughly agrees with the peak injection energies (100-300meV) seen in Fig.~\ref{fig:asym}(b).
We note that the numerical data have the minimum above the Fermi level while the STM spectra have a V-shape gap with the minimum at zero bias.
The V-shape gap might be attributed to the soft Coulomb gap \cite{es75} or soft Hubbard gap \cite{si09} caused by an interplay of electron correlations and randomness, as indicated by the strong charge inhomogeneity observed by the STM at surfaces.\cite{hl04}

ARPES allows a more detailed comparison of the spectra.
Yang {\it et al.} \cite{yr08} succeeded in deriving low-energy ($\w\lesssim 0.03$eV) spectra above the Fermi level by carefully analyzing the ARPES data for Bi$_2$Sr$_2$CaCu$_2$O$_{8+\d}$. 
This was done by using the fact that ARPES data at finite temperatures contains information of unoccupied states because of the smeared tail of the Fermi distribution function for $\w > 0$. 
In the superconducting state they observed electron-hole symmetric spectra both in the nodal and antinodal region, in accordance with the BCS spectral function.\cite{bc57}
Meanwhile in the pseudogap state they found that 
(i) around the antinode the spectrum is nearly symmetric 
with intense peaks below and above $E_{\text{F}}$ separated by a gap of $\sim 0.06$eV,
and that (ii) as approaching the node the peak below $E_{\text{F}}$ goes up and eventually crosses $E_{\text{F}}$ while the peak above $E_{\text{F}}$ disappears from the measured energy range.
While (ii) clearly shows the asymmetry as to $E_{\text{F}}$,
Yang {\it et al.} \cite{yr08} interpreted (i) as an evidence of preformed pairing.

Liebsch and Tong \cite{lt09} obtained an asymmetry similar to (ii) with the 
CDMFT.
Nevertheless it is still worthwhile to see if the symmetry (i) can be reproduced since it is relevant to the mechanism of the pseudogap.
Figure \ref{fig:asym}(c) shows the normal-state spectral function $A(\Vec{k},\w)$, calculated with the CDMFT+ED using $\eta=0.03t$.
We find that around the antinode, the two intense peaks reside nearly symmetric with opening of a gap. 
As approaching to nodal region, the gap below the Fermi level monotonically decreases and vanishes when the lower peak reaches the Fermi level [right panel of Fig.~\ref{fig:asym}(c); see also Fig.~\ref{fig:fig2}(a)].
These behaviors are consistent with the experimental observations (i) and (ii).
We note that in the $t'=0$ case [Fig.~\ref{fig:fig2}(c)] the gap at $(\pi,0)$ above $E_{\text{F}}$ is larger than that below $E_{\text{F}}$, 
whereas the spectrum is more symmetric for $t'=-0.4t$ [Fig.~\ref{fig:asym}(c)], which is more appropriate for Bi$_2$Sr$_2$CaCu$_2$O$_{8+\d}$.

In the nodal direction, the gap lies above $\w\simeq 0.1t$.
This is again consistent with the ARPES \cite{yr08} which saw the region below $0.03\text{eV} \sim 0.08t$ and observed no gap in this direction.
We note that this is distinct from the picture in Ref.~\onlinecite{kr06}, where  the gap was assumed to close in the nodal direction.

Thus we have shown that the ARPES data \cite{yr08} does not necessarily indicate a preformed pairing, but is rather naturally interpreted as a consequence of the dispersive zero surface.

\subsection{Back-bending behavior of dispersion}\label{ssec:back}

\begin{figure}[t]
\center{
\includegraphics[width=0.48\textwidth]{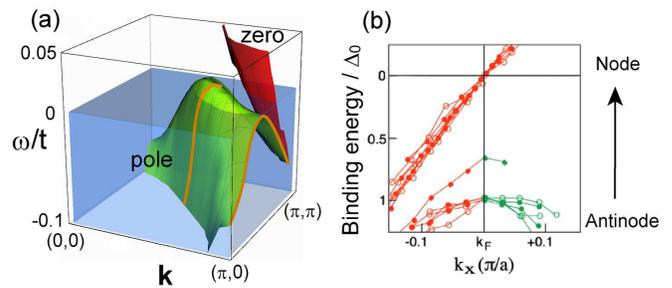}}
\caption{(color online). 
       (a) Enlarged view of Fig.~\ref{fig:fig2}(a) at low
       energy. Solid (orange) curves are guides to the eye for 
       band dispersions in nodal and antinodal regions.
       (b) Binding energy of the ARPES spectral peaks 
       along momentum cut with regular $k_y$ intervals. Reproduced
       from Fig.~2(d) in Ref.~\onlinecite{kc08}.
       The data was taken for the pseudogap state ($T=140$K) of 
       a thin film Bi$_2$Sr$_2$CaCu$_2$O$_8$ sample with 
       $T_{\text{c}}=90$K.
       $\D_{\text{0}} (\sim 50 \text{meV})$ is the maximum of 
       the $d$-wave gap fitting the data.
       }
\label{fig:back}
\end{figure}

Another important aspect captured in Fig.~\ref{fig:fig2}(a) is 
the back-bending behavior of the band cutting the Fermi level.
Namely, the pole surface, which monotonically goes up to $(\pi,\pi)$ in the bare dispersion, is bent back below $E_{\text{F}}$ around $(\pi,\pi)$.
This is more clearly seen in Fig.~\ref{fig:back}(a), which is an enlarged view of Fig.~\ref{fig:fig2}(a) in a lower energy range. 
The back bending can be seen below $E_{\text{F}}$ around the antinodes while it is above $E_{\text{F}}$ around the nodal direction.

Actually the ARPES \cite{kc08} observed a similar behavior.
Figure \ref{fig:back}(b) is a reproduction of Fig.~2(d) in Ref.~\onlinecite{kc08}, where the back-bending behavior is seen only around the antinode.
In the light of Fig.~\ref{fig:back}(a) the absence of the back bending around the node will be simply because the top of the band is located at a higher energy than the maximum energy of the measurement.
Note that $t'<0$ lifts (lowers) the band around the nodes (antinodes).  
Enhancing $-t'$ shifts the Fermi surface to $(0,0)$ as well, as we will discuss in Sec.~\ref{ssec:tp}.
Then a better agreement with the ARPES data may be reached. 

In Ref.~\onlinecite{kc08} the back-bending behavior around the antinodes was interpreted as an evidence of preformed Cooper pairs in the pseudogap state because it resembles a band dispersion in the BCS superconductors.\cite{bc57}
However, our result implies another simple interpretation: The band is pushed down by the neighboring zero surface, which cuts the Fermi level around $(\pi,\pi)$, due to the large self-energy around it.
This picture is confirmed in Fig.~\ref{fig:fig2}(d), where we plot
the energy dependence of the self-energy at the Fermi momentum $\Vec{k}_{\text{F2}}$ closest to the zero surface.
We see that the real part of the self-energy is negatively large below the zero at $\w\simeq0.06t$, which should push down the band around the momentum.
The result indicates that the pair formation is not necessary and 
the zeros of $G$ resulting directly from strong correlation effects
give an alternative interpretation of the back-bending dispersions.

Recently, an evidence against the scenario that the back bending is a consequence of preformed pair was further reported.\cite{hh10}
In this ARPES observation, the back-bending momentum, i.e., the location of the pole at the lowest binding energy along momentum cuts, is clearly deviated from the Fermi momentum at above the pseudogap opening temperature.
In contrast, we note that the two momenta should agree in the BCS theory and in the preformed pair scenario as well.
Furthermore, the ARPES reported that the back-bending momentum along $(\pi,0)$-$(\pi,\pi)$ shifts closer to $(\pi,\pi)$ than the Fermi momentum.
This indicates a zero surface around $(\pi,\pi)$ at the Fermi level and a center of the gap residing above $E_\text{F}$, in full consistency with our pole-zero structure Fig.2(a).

We note that a back-bending dispersion was already observed in an early QMC study,\cite{ph97} where antiferromagnetic fluctuations were proposed as the origin of the pseudogap.
Although our numerical data do not exclude the antiferromagnetic fluctuations from the possible mechanisms, the less-$\Vec{k}$-dependent pseudogap as well as the asymmetric location of the hole pocket (see Sec.~\ref{ssec:arc}) seems to oppose the mechanism; instead it rather supports that the pseudogap is a direct consequence of the proximity to the Mott insulator.

\subsection{Energy-distribution curve}\label{ssec:edc}

\begin{figure}[t]
\center{
\includegraphics[width=0.48\textwidth]{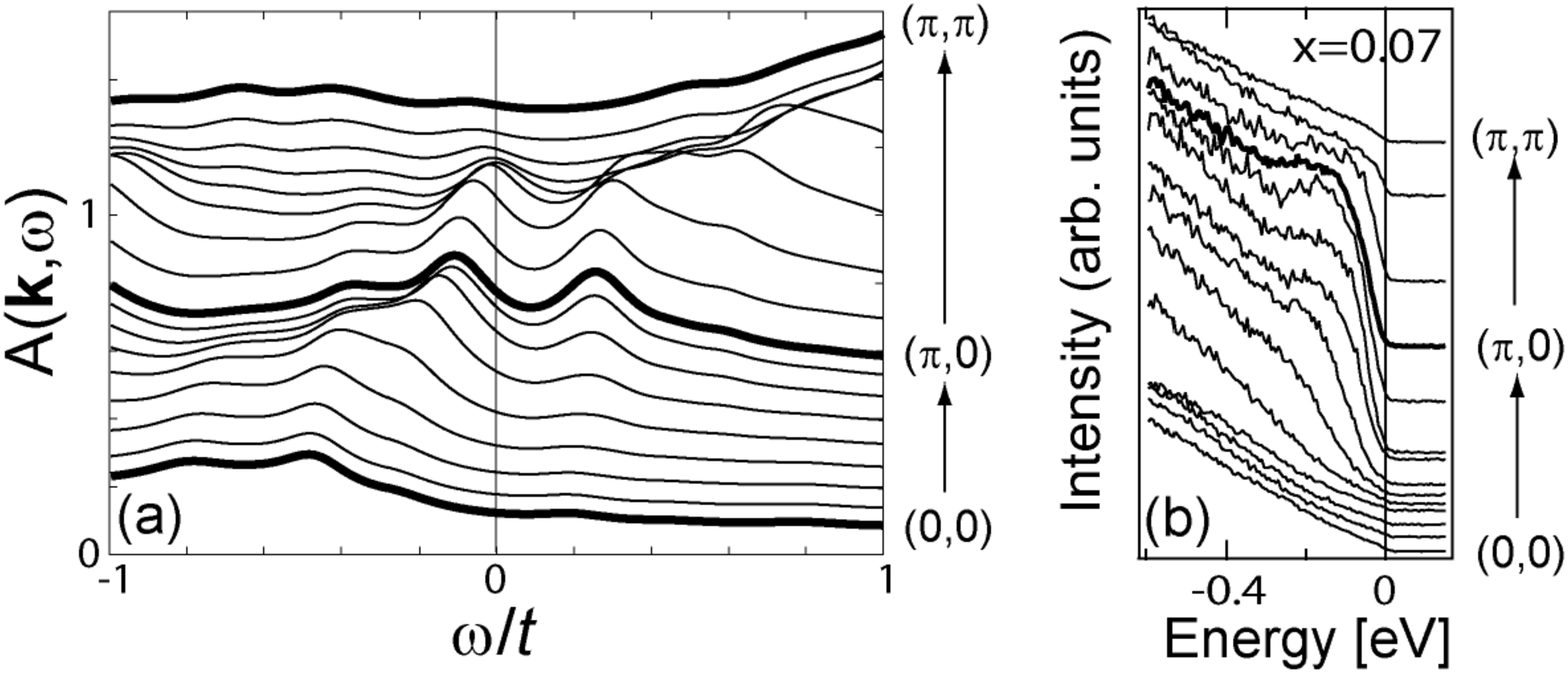}}
\caption{(a) Energy-distribution curves calculated with the CDMFT+ED 
       along symmetric lines $(0,0)\text{-}(\pi,0)\text{-}(\pi,\pi)$ 
       for $t'=-0.2t$, $U=12t$, $n=0.93$, and $\eta=0.1t$.
       For clarity the curves are offset 
       by 0.06 (0.1) for $(0,0)\text{-}(\pi,0)$ [$(\pi,0)\text{-}(\pi,\pi)$].
       (b) ARPES spectra for La$_{2-x}$Sr$_x$CuO$_4$ ($x=0.07$), 
       reproduced from Fig.~5 in Ref.~\onlinecite{yz07}.
       }
\label{fig:edc}
\end{figure}

Zeros of $G$ are not directly seen in spectra.
Their footprints may, however, be detected through a sudden suppression of spectra due to a large Im$\S$ around the zeros.
Figure \ref{fig:edc}(a) shows energy-distribution curves (EDC) of the spectral function along momentum cuts $(0,0)\text{-}(\pi,0)\text{-}(\pi,\pi)$, calculated by the CDMFT+ED with $\eta=0.1t$ for $t'=-0.2t$, $U=12t$, and $n=0.93$.
The results exhibit a coherent peak around $(\pi,0)$ just below the Fermi level, its shift to lower energy from $(\pi,0)$ to $(\frac{\pi}{2},0)$,
and the incoherent feature around $(0,0)$ and $(\pi,\pi)$. 
All these features are consistent with ARPES data [Fig.~5 in Ref.~\onlinecite{yz07}, reproduced in Fig.~\ref{fig:edc}(b)].
This agreement supports our zero mechanism and shows that 
the incoherent feature can be interpreted as the effect of zeros of $G$: 
Since the zero surface exists just above the band around $(\pi,\pi)$ [Fig.~\ref{fig:fig2}(a)] and since another one is located just below the band around $(0,0)$ [not shown in Fig.~\ref{fig:fig2}(a), but can refer to Fig.~\ref{fig:rkw_tp}(b) below], 
the spectrum associated with the band is smeared around these momenta due to the large Im$\S$ around the zeros.
While the suppression around $(\pi,\pi)$ is relevant to the emergence of Fermi arc,\cite{sk06,sm09,sm09-2} that around $(0,0)$ is related to the waterfall behavior discussed in the next subsection.

\subsection{Waterfall}\label{ssec:wf}
\begin{figure}[t]
\center{
\includegraphics[width=0.48\textwidth]{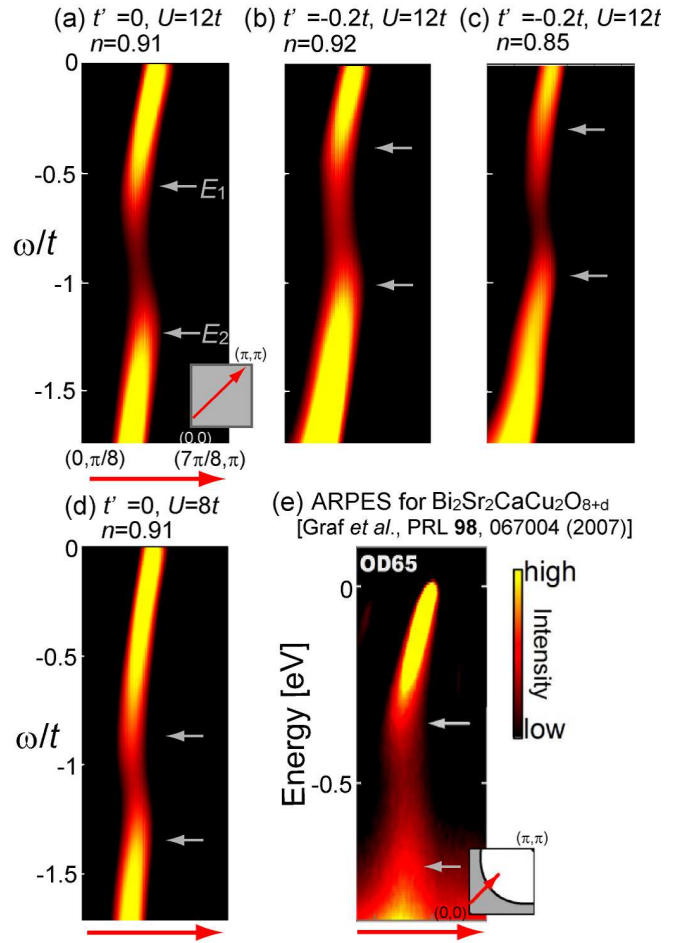}}
\caption{(color online). 
       (a)-(d) Spectral distributions 
       calculated with the CDMFT+ED along the momentum cut 
       $(0,\frac{\pi}{8})\text{-}(\frac{7\pi}{8},\pi)$ 
       [denoted by a red arrow in the inset of (a)] 
       for a variety of parameter sets.
       The data are broadened with $\eta=0.2t$.
       Grey arrows denote the energies, $E_1$ and $E_2$, 
       discussed in the text.
       (e) ARPES spectrum for overdoped Bi$_2$Sr$_2$CaCu$_2$O$_{8+\d}$,
       reproduced from Fig.~1(c) in Ref.~\onlinecite{gg07}. 
       }
\label{fig:wf}
\end{figure}

Recent ARPES \cite{kb05,gg07,xy07,vk07,mz07} observed an anomalous spectral structure, called ``waterfall", at high binding energies $\sim 0.3\text{-}0.4$eV, 
where the band cutting the Fermi level suddenly loses the spectral weight and starts falling down to $\sim -0.7$eV with a suppressed intensity.
Below the energy a strong intensity emerges again from nearly the same momentum as that the waterfall starts.
This high-energy anomaly has been found quite generally in hole-doped 
cuprates, irrespective of the presence or absence of superconductivity, 
under, optimally or overdoped, and detailed compositions.
Moreover similar features have been reported in other transition-metal compounds such as SVO$_3$ \cite{yt05} and LaNiO$_3$.\cite{ec09}
These suggest a universality of the phenomenon in strongly-correlated metals.\cite{bk07}

The dynamical cluster approximation (DCA) + QMC study \cite{mj07} for the 2D Hubbard model with $t'=0$ reported a similar structure in the spectra.
Based on the similarity of the DCA results to those with a perturbative calculation incorporating antiferromagnetic spin fluctuations, the authors proposed that the waterfall results from high-energy spin fluctuations.
Here we implement a CDMFT+ED study, and closely examine how the pole-zero structure underlies the waterfall phenomenon and 
how the energy scale of the waterfall depends on model parameters. 
Our analysis indicates alternative interpretation for the phenomenon.

Figures \ref{fig:wf}(a)-(d) plot the spectral intensity calculated along the momentum cut from $(0,\frac{\pi}{8})$ to $(\frac{7\pi}{8},\pi)$ for various parameter sets.
The results show nice resemblances with the ARPES result [Fig.~1(c) in Ref.~\onlinecite{gg07}, reproduced in Fig.~\ref{fig:wf}(e)], i.e., 
(i) an abrupt change in the slope of the band around $\w=E_1$, accompanied by the simultaneous reduction of the intensity,
(ii) nearly vertical dispersion with suppressed intensity in the waterfall region, and
(iii) reemergence of a strong intensity around $\w=E_2$.
The energy scale also agrees roughly with the experimental value, if one keeps in mind $t\sim 0.4$ eV.

We investigate how $E_1$ and $E_2$ depend on $t'$, $n$, and $U$.
First, comparing Figs.~\ref{fig:wf}(a) and (b), we see that $t'<0$ reduces both $|E_1|$ and $|E_2|$. 
Second, comparison of Figs.~\ref{fig:wf}(b) and (c) shows that doping also reduces the energies. 
This doping dependence is qualitatively consistent with the DCA results \cite{mj07} for 16 sites cluster at a finite temperature.
Third, the decrease of $U$ from $12t$ [Fig.~\ref{fig:wf}(a)] to $8t$ [Fig.~\ref{fig:wf}(d)] increases the energies.

These parameter dependences can be naturally explained as follows.
In general, strong correlation effects make the width of coherent band
(i.e., the band above $E_1$) 
narrower by increasing the mass of the low-energy particles.
This leads to the decrease of $|E_1|$ as $U$ increases.
Indeed the slope of the coherent band decreases from $1.3ta$ ($a$: the lattice constant) to $1.0ta$ as $U$ increases from $8t$ to $12t$.
Meanwhile $t'$ lifts the band around the nodal direction, 
as will be discussed in Sec.~\ref{ssec:tp}.
This elevates $E_1$ and $E_2$.
The doping dependence seen in Figs.~\ref{fig:wf}(b) and (c) can be qualitatively understood as a downward shift of the chemical potential with doping.
As we showed in Ref.~\onlinecite{sm09}, in hole-doped Mott insulators doping causes a rigid-band-like shift of the chemical potential, though it cannot be described within the single-electron picture. 
In fact, the low-doping phase is 
not adiabatically connected with the Fermi liquid phase at higher dopings because of the intervening Lifshitz transition and the zero-surface emergence.

The above interpretation suggests that the energy scale of the waterfall is not necessarily related to the antiferromagnetic spin 
fluctuations \cite{mj07} but is considered to be a rather direct consequence of electron correlation, i.e., the mass renormalization of the coherent band.
The picture is corroborated by the fact that a similar behavior can be seen within the single-site DMFT.\cite{bk07}

In view of the pole-zero structure in Fig.~1(b) in Ref.~\onlinecite{sm09}, the waterfall emerges in the energy region $-2t\lesssim\w\lesssim-t$, where many smeared pole and zero surfaces pile.
The congestion of poles and zeros within the finite-cluster calculation indicates an incoherent nature of this energy-momentum region, which would result in the waterfall.
Notice that the waterfall obtained in this paper has neither a gap nor dispersive features \cite{mj07}, but a vertical structure with a suppressed intensity. 
This unusual structure comes out thanks to the use of the cumulant periodization (see Sec.~\ref{sec:method}) which can describe the strong momentum dependence of the self-energy.
Below the pile we see a relatively coherent pole surface extending down to $\w\sim -3.5t$. This corresponds to the band reemerging at the high binding energy [(iii)].
Note that a comparison at higher binding energies is difficult because the ARPES spectra are overlapped by other Cu-$d$ or O-$p$ bands.

\subsection{Low-energy kink in dispersion}\label{ssec:kink}

\begin{figure}[t]
\center{
\includegraphics[width=0.35\textwidth]{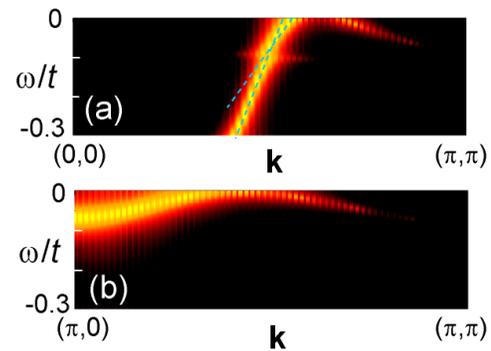}}
\caption{(color online). Spectral function calculated with the CDMFT+ED 
        for $t'=0$, $U=8t$, $n=0.93$, and $\eta=0.005t$ along 
        the momentum cuts
        (a) $(0,0)\text{-}(\pi,\pi)$ and (b) $(\pi,0)\text{-}(\pi,\pi)$.
        The (blue) dashed lines in (a) are guides to the eye.
       }
\label{fig:kink}
\end{figure}

In prior to the discovery of the waterfall, another anomaly (kink) in dispersion has been found in ARPES at lower binding energies \cite{bl00} and its origin has been in dispute.
The kink is located at a binding energy around $0.05$eV, where the band cutting the Fermi level sharply changes its slope.
In the normal state the kink is clearly seen around the nodal region while, as approaching the antinodal region, the band becomes flatter and then the kink becomes less visible.
The band structure continuously changes from the nodal to antinodal region where a pseudogap exists just above the band.\cite{kr01,sm03}

The CDMFT+ED results in Fig.~\ref{fig:kink} show similar behaviors:
(i) The band suddenly changes the slope at a kink around $\w=-0.1t\sim -0.04$eV
in the nodal region [Fig.~\ref{fig:kink}(a)],
and (ii) as approaching the antinode, the band becomes flatter and the 
slope change becomes weaker [Fig.~\ref{fig:kink}(b)].

In general, the zero surface which generates the pseudogap pushes down the dispersion near the Fermi level. 
This makes a quick change of the slope of the dispersion distinct from the part deeply below the Fermi level, which may be the underlying origin of the kink formation.
However, the sudden change of the dispersion observed in Fig.~\ref{fig:kink}(a) naturally requires a precursory formation of a tiny zero surface around $\w=-0.1t$.
This means a coupling of the quasiparticle with some other excitations with this energy, whichever bosonic or fermionic.
The energy resolution of the present cluster size is not obvious and
it could be an artifact of the present small cluster calculation.
Nevertheless, it is remarkable to note that the observed kink in Fig.~\ref{fig:kink}(a) is rather universal and we see similar kinks in other part of the quasiparticle dispersions, for instance in Fig.~\ref{fig:rkw_tp}(b) below. 
This may be alternatively interpreted that, in the strongly correlated region, the quasiparticle is strongly renormalized and may couple to various intrinsic electronic modes of charge and spin origins.

In high-$T_{\text{c}}$ cuprates the mechanism of the kink has been extensively discussed in the literature from the viewpoint of a coupling of electrons with some bosonic mode, such as phonon \cite{lb01} and magnetic ones.\cite{jv01,kr01}
Our result, however, implies that the kink may be a rather general phenomenon in the proximity to the Mott insulator,\cite{bk07} while the specific mechanism of kink formation is left for future studies.

\subsection{Fermi arc and hole-pocket Fermi surface}\label{ssec:arc}

Next we shift our focus on the zero-energy electronic structure, namely, Fermi surface.
The spectra in hole-doped cuprates have been extensively studied by ARPES.\cite{dh03}
One of the most remarkable findings in these studies is the observation of truncated Fermi surfaces, called ``Fermi arc".\cite{nd98}

\begin{figure}[t]
\center{
\includegraphics[width=0.4\textwidth]{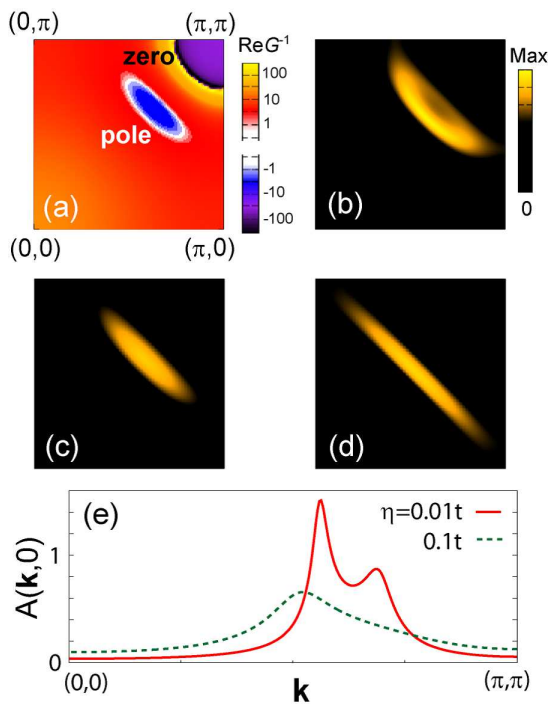}}
\caption{(color online). Momentum maps of (a) Re$G^{-1}(\Vec{k},\w=0)$ 
       and (b) $A(\Vec{k},\w=0)$ 
       calculated with the CDMFT+ED for $t'=0$, $U=12t$, and $n=0.91$.
       $\eta=0.0001t$ $(0.01t)$ is used for (a) [(b)].
       CDMFT+CTQMC results of $-\beta G(\Vec{k},\frac{\b}{2})$ for 
       $t'=0$, $U=12t$ and $T=0.025t$ at (c) $n=0.95$ and (d) $n=0.90$.
       (e) Momentum-distribution curves calculated with the CDMFT+ED 
       along $(0,0)$-$(\pi,\pi)$ for the same parameters as 
       those for (a) and (b).
       $\eta=0.01t$ (solid curve) and $0.1t$ (dashed curve).
       }
\label{fig:pocket}
\end{figure}

Several authors have discussed the Fermi arc in terms of coexisting poles and zeros of $G$.\cite{sm09,kr06,et02,bg06,ks06,sk06,sm09-2,sp07}
In this mechanism hole-pocket Fermi surfaces around the nodal directions coexist with a zero surface around $(\pi,\pi)$.
Because Im$\S$ is large around the zero surface, the pockets lose the spectral intensity more on the side closer to the zero surface, leaving arc-like spectra on the opposite side.
This is reproduced in Figs.~\ref{fig:pocket}(a) and (b).
Figure \ref{fig:pocket}(a) shows the underlying pole-zero structure calculated 
by the CDMFT+ED at $T=0$ without any substantial smearing to the singularities, where a clear hole-pocket Fermi surface coexists with a zero surface.
When we introduce a smearing to the singularities (see Sec.~\ref{ssec:ed}), we obtain the spectral map, Fig.~\ref{fig:pocket}(b),
in which the arc-like structure is observed.

We confirm the zero mechanism by directly calculating a spectral weight at finite temperatures by employing the CTQMC method as the solver for the CDMFT.
To obtain the spectrum without through any analytic continuation procedure which inevitably suffers from a large error bar, we calculate  
\begin{align}\label{eq:ghb}
-\b G(\Vec{k},\t=\beta/2)=\frac{\b}{2}
\int_{-\infty}^{\infty}\frac{A(\Vec{k},\w)}{\cosh(\b\w/2)}d\w
\end{align}
as a function of $\Vec{k}$.
This quantity is an integral of the spectral weight over a width $\sim T$ around $\w=0$ and approaches $A(\Vec{k},\w=0)$ 
for large $\beta$, so that gives an estimate of 
$A(\Vec{k},\w=0)$ at low temperatures.\cite{footnote2}
In Figs.~\ref{fig:pocket}(c) and (d) we plot the results at $T=0.025t$.
For $n=0.95$ [Fig.~\ref{fig:pocket}(c)] we see an arc at a location similar to the one in Fig.~\ref{fig:pocket}(b) while for $n=0.9$ [Fig.~\ref{fig:pocket}(d)] the spectra extend to the antinodal regions. 
The doping evolution of the spectra is qualitatively consistent with that obtained with the CDMFT+ED in Fig.~2(c)-(e) in Ref.~\onlinecite{sm09}.
The qualitative agreement between the CDMFT+ED result with broadening $\eta$ and the CDMFT+CTQMC result at finite temperatures supports that the phenomenological smearing factor $\eta$ well simulates the thermal effects. 
Moreover it corroborates the above-mentioned zero mechanism for the emergence of the Fermi arc.
We note that the zero surface around $(\pi,\pi)$ is also consistent with
the results by other cluster schemes such as DCA,
which observed a strong scattering amplitude in the momentum patch
around $(\pi,\pi)$ in both $N_\text{c}=4$ (Ref.~\onlinecite{gw08-2}) and
$N_\text{c}=8$ (Ref.~\onlinecite{wg09}) calculations. 

Interestingly, recent high-resolution ARPES \cite{ml09} reported the existence of hole-pocket Fermi surfaces around the nodal directions in underdoped Bi$_2$Sr$_{2-x}$La$_x$CuO$_{6+\d}$.
The observed hole pockets have much less intensity on the side 
closer to $(\pi,\pi)$ than the opposite.
This is consistent with the zero mechanism in Fig.~\ref{fig:pocket}(b).
Important findings in the ARPES are that (i) the hole pockets are not located symmetrically with respect to the antiferromagnetic Brillouin zone boundary, i.e., $(\pi,0)\text{-}(0,\pi)$ line (and its symmetrically-related ones), and that (ii) the spectral intensity on the $(\pi,\pi)$ side is finite even in the nodal directions.
(i) excludes several scenarios for the pseudogap, which attribute the pockets to symmetry breakings \cite{ks90,v97,cl01} because in these theories hole pockets should be centered symmetrically as to the $(\pi,0)\text{-}(0,\pi)$ line.\cite{ml09} 
Meanwhile (i) is consistent with our zero mechanism that does not assume any symmetry breaking.
(ii) is at odds with the theory based on an ansatz
that gives a zero intensity in the outer point of the pocket crossing the nodal direction,\cite{ml09} while it agrees well with our numerical data in Fig.~\ref{fig:pocket}(b).
This is more clearly seen in Fig.~\ref{fig:pocket}(e), where we plot $A(\Vec{k},0)$ along the momentum cut in the nodal direction:
For $\eta=0.01t$, in addition to the main peak at $\Vec{k}\simeq(0.56\pi,0.56\pi)$, we see the secondary peak at $\simeq (0.7\pi,0.7\pi)$, corresponding to the hole pocket structure.
The second peak is, however, not visible for $\eta=0.1t$ and only the broadened main peak can be seen there.
Namely, in our theory, the spectrum looks either a pocket like or an arc like 
depending on how the incoherence due to the zero surface is strong.
The strength of the incoherence is controlled by the value of $\eta$ or temperature, or the distance between zeros and poles, as partly demonstrated in Fig.~4 in Ref.~\onlinecite{sm09}.
Figure \ref{fig:pocket}(e) also suggests that the energy resolution required to detect the pocket in the example of the pole-zero structure as Fig.~\ref{fig:pocket}(a) is in between $0.01t(\sim 4\text{meV})$ and $0.1t(\sim 40\text{meV})$, which roughly corresponds to the energy resolution in ARPES.
This consideration on the incoherence explains why the pocket had not been detected until the recent high-resolution ARPES.\cite{ml09}

One obvious difference between Fig.~\ref{fig:pocket}(b) and the ARPES \cite{ml09} is the location of the pocket: The pocket in Fig.~\ref{fig:pocket}(b) resides closer to $(\pi,\pi)$ than that observed in the ARPES.
This may be attributed to the difference between the model parameters we used and realistic ones for Bi$_2$Sr$_{2-x}$La$_x$CuO$_{6+\d}$.
In particular, we show in the next subsection that the next-nearest-neighbor transfer $t'$, which is zero in Fig.~\ref{fig:pocket}(b), indeed shifts the pocket in the direction to $(0,0)$.

\subsection{Effect of the next nearest-neighbor transfer $t'$}\label{ssec:tp}

\begin{figure}[t]
\center{
\includegraphics[width=0.48\textwidth]{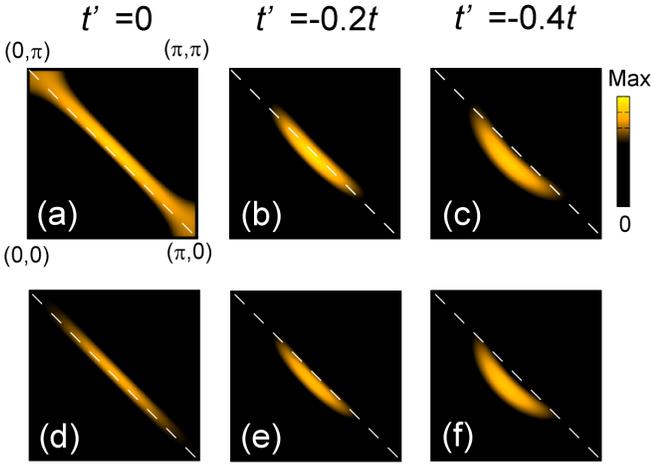}}
\caption{(color online). $t'$ dependence of $-\beta G(\Vec{k},\frac{\b}{2})$ 
       in Eq.~(\ref{eq:ghb}) at $T=0.05t$ for 
       (a)(b)(c) $U=8t$ and $n=0.93$, and (d)(e)(f) $U=12t$ and $n=0.90$. 
       White dashed line represents the antiferromagnetic 
       Brillouin zone boundary.
      }
\label{fig:arc_tp}
\end{figure}

Here we systematically study the effect of $t'$ on the electronic structure in lightly hole- and electron-doped regions.
Figure \ref{fig:arc_tp} shows $t'$ dependence of the integrated spectra, Eq.~(\ref{eq:ghb}), calculated with CDMFT+CTQMC at $T=0.05t$ for (a)-(c) $U=8t$ and $n=0.93$ and (d)-(f) $U=12t$ and $n=0.90$.
We see that the arc shifts to $(0,0)$ as $-t'/t$ increases, and that at $t'=-0.4t$, which is a reasonable value for Bi$_2$Sr$_{2-x}$La$_x$CuO$_{6+\d}$, the arc resides inside of the antiferromagnetic Brillouin zone, in consistency with experiments.\cite{ml09}
We also see that $t'$ increases the curvature of the arc.
A comparison with ARPES 
in further detail will require a more realistic determination of the model parameters, longer-range transfer integrals and Coulomb interactions, and a calculation on a larger cluster.
These remain for future researches.

\begin{figure}[t]
\center{
\includegraphics[width=0.48\textwidth]{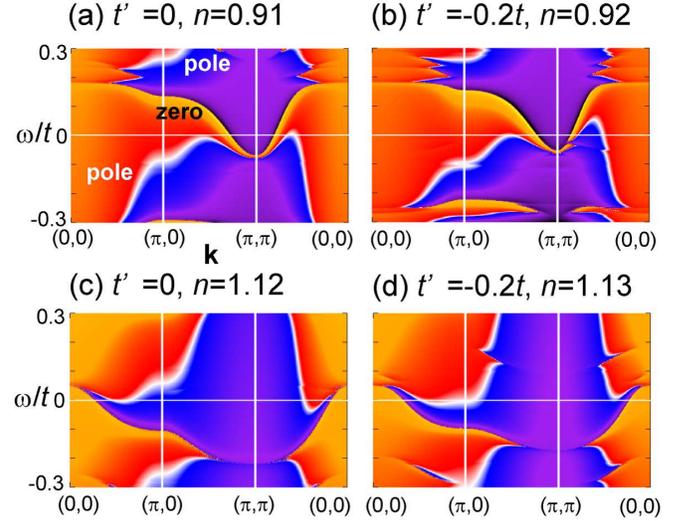}}
\caption{(color). Re$G^{-1}(\Vec{k},\w)$ calculated with the CDMFT+ED
       along symmetric lines for $U=12t$.
       (a) $t'=0$ and $n=0.91$,
       (b) $t'=-0.2t$ and $n=0.92$, 
       (c) $t'=0$ and $n=1.12$, and
       (d) $t'=-0.2t$ and $n=1.13$. 
       $\eta=0.001t$ is used.
       Color scale is the same as that in Fig.~\ref{fig:fig2}(c).
      }
\label{fig:rkw_tp}
\end{figure}

The above shift of the arc with $t'$ can be understood by plotting the underlying pole-zero structures.
Figures \ref{fig:rkw_tp}(a) and (b) compare the structures for $t'=0$ and $t'=-0.2t$ in a lightly hole-doped region.
The figures show that at low energies $t'$ lifts the poles in the nodal direction while lowers them around the antinodes, as already expected from the bare dispersion Eq.~(\ref{eq:disp}).
Then the hole pocket expands and the Fermi surface with the stronger intensity shifts to $(0,0)$ direction.
We note that, as we discuss in the next subsection, the third-neighbor transfer integral, which is not taken into account in the present calculation but exists in real materials, further enhances the pocket around $(\frac{\pi}{2},\frac{\pi}{2})$.

The effect of $t'$ is totally different for electron-doped cases.
Figures \ref{fig:rkw_tp}(c) and (d) show the structures for $t'=0$ and $-0.2t$, respectively, in an electron-doped region.
The effect of $t'$ at low energies is again understood by Eq.~(\ref{eq:disp}):
$t'$ lifts the poles in the nodal direction while lowers around the antinodes.
However, the consequence on the Fermi surface completely differs from that in hole-doped cases.
For the particle-hole symmetric $t'=0$ case, 
the electron pockets appear around the nodes for electron doping, 
corresponding to the hole pockets in hole-doped cases, 
however, for finite $t'$, 
the electron pockets appear around the antinodes,
as shown in Fig.~\ref{fig:rkw_tp}(d). 
This is because the low-energy zero surface pushes up the dispersion around $(0,\frac{\pi}{2})$. 
We compare the Fermi surface structure with ARPES for the electron-doped cuprates in detail in Sec.~\ref{ssec:ele}.

\subsection{High-energy spectra in electron-doped cuprates}\label{ssec:ele2}

\begin{figure}[t]
\center{
\includegraphics[width=0.48\textwidth]{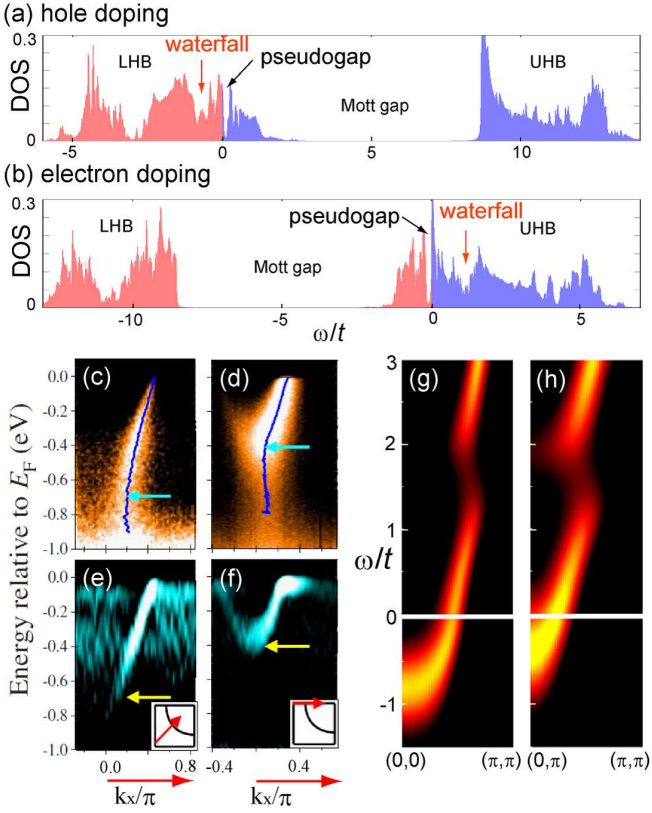}}
\caption{(color online). 
       (a)[(b)] DOS for $t'=-0.2t$, $U=12t$ and $n=0.92$($1.09$) 
       with $\eta=0.01t$.
       LHB and UHB represent the lower and upper Hubbard bands, respectively.
       (c)(d) ARPES spectra and (e)(f) their second derivatives in $\w$
       for Nd$_{1.85}$Ce$_{0.15}$CuO$_4$. Reproduced from Fig.~1 
       in Ref.~\onlinecite{iy09}.
       (g)(h) Spectra calculated with the CDMFT+ED for $t'=-0.4t$, $U=12t$,
       and $n=1.11$ along momentum cuts, $(0,0)\text{-}(\pi,\pi)$ and 
       $(0,\pi)\text{-}(\pi,\pi)$, respectively.
       $\eta=0.3t$ is used.
       }
\label{fig:wf2}
\end{figure}

Recently high-energy spectra of an electron-doped cuprate Nd$_{1.85}$Ce$_{0.15}$CuO$_4$ were studied by ARPES.\cite{iy09}
To discuss the results we first emphasize that here the ARPES observes a totally different energy region from that for hole-doped cuprates.
This is illustrated in Figs.~\ref{fig:wf2}(a) and (b):
ARPES observes the lower Hubbard band (LHB) in hole-doped cases while in electron-doped cases it does ingap states between the Mott gap and the pseudogap, in addition to the pseudogap structure itself.
As mentioned in Sec.~\ref{ssec:eh}, with the electron-hole transformation the ARPES result can be interpreted as that for unoccupied states in a hole-doped system.
Therefore the results in Ref.~\onlinecite{iy09} provide a precious opportunity to compare the calculated spectra of ingap states with experiments.
The pseudogap and the Fermi surface are discussed in the next section.

The ARPES spectra and its second derivative in $\w$ (Fig.~1 in Ref.~\onlinecite{iy09}) are reproduced in Figs.~\ref{fig:wf2}(c)(d) and (e)(f), respectively.
Around the antinode the spectra [Fig.~\ref{fig:wf2}(d)] suddenly lose the intensity at $\w \simeq -0.4$eV.
The analysis on the second derivative [Fig.~\ref{fig:wf2}(f)] shows that the low-energy band reaches the bottom at this energy.
On the other hand, around the node the low-energy band persists down to $\w \simeq -0.7$eV [Fig.~\ref{fig:wf2}(c)(e)].

Figures \ref{fig:wf2}(g) and (h) depict the spectra calculated with the CDMFT+ED for $t'=-0.4t$, $U=12t$, and $n=1.11$ along momentum cuts, $(0,0)\text{-}(\pi,\pi)$ and $(0,\pi)\text{-}(\pi,\pi)$, respectively.
For $\w<0$ we see that the band bottom is located at $\w\simeq -0.7t$ around $(0,\pi)$ while it is located at a higher binding energy $\w\simeq -t$ around $(0,0)$.
This is consistent with the ARPES data.
Note that the numerical data do not show any intensity below the band bottoms, $\w < -0.7t$ ($-t$) around $(0,\pi)$ [$(0,0)$], which is within the gap region between LHB and the ingap states, 
whereas the ARPES spectra show waterfall-like structures in $-1\text{eV}\lesssim \w \lesssim -0.7\text{eV}$ ($\w \lesssim -0.4\text{eV}$) around $(0,\pi)$ [$(0,0)$]. 
The origin of this discrepancy is not clear, but 
one possibility is a multiband effect, which is not considered in the present single-band model. 

Meanwhile for $\w>0$, we see a waterfall in the dispersion at $\w\simeq t\text{-}2t$.
This corresponds, with the electron-hole transformation, to the waterfall in hole-doped cases.
The result predicts that this anomaly will be observed in electron-doped cuprates when the unoccupied spectra become available up to the high energy.\cite{footnote5}

\subsection{Pseudogap and Fermi pockets in electron-doped cuprates}\label{ssec:ele}

Lastly we discuss low-energy electronic structures in electron-doped cases.
The pseudogap has been experimentally observed also in electron-doped cuprates,\cite{ot01} although the doping range is much more limited than that in hole-doped cases due to the wider antiferromagnetic region near the undoped Mott insulator.
The mechanism of the pseudogap has extensively been discussed within weak-coupling theories \cite{kr03,kh04}, which has successfully reproduced various experimental results such as the doping evolution of the Fermi surface.
In these theories the mechanism has been ascribed to the antiferromagnetic long-range correlations.
On the other hand, in the strong-coupling regime the cluster-perturbation theory \cite{st04} and the CDMFT \cite{cc05,kk06} found that the pseudogap results from nonlocal but short-ranged dynamics without any long-range order or correlations.
As pointed out in Ref.~\onlinecite{st04} and \onlinecite{kk06}, the gap amplitude does not scale as $J\sim t^2/U$, so that the short-ranged dynamics is not simply ascribed to the AF correlation. 
Because it is unknown which mechanism, weak-coupling or strong-coupling, is relevant to electron-doped cuprates,
it is worthwhile to see whether and how experimental data are understood in 
the strong-coupling picture, especially from the perspective of zeros of $G$.

The available ARPES data for the electron doped cuprates seems to have relatively poor resolutions and a clear dispersion has not been reported. Nevertheless, a recent ARPES~\cite{iy09-2} measurement appears to have revealed the structure of the pseudogap for a family of electron-doped cuprates, $Ln_{2-x}$Ce$_x$CuO$_4$ ($Ln=$ Nd, Sm, and Eu).
In Fig.~2 in Ref.~\onlinecite{iy09-2} the EDC peak position jumps from $\w\sim-0.03$eV to $\w\sim-0.15$eV around antinode, which implies the presence of the pseudogap opening below the electron pocket.
This highly asymmetric position of the pseudogap is a feature which cannot be seen by ARPES for the hole-doped cuprates but observable for electron-doped ones: Because in electron-doped cases the zero surface forming the pseudogap mainly extends below $E_\text{F}$ [see Fig.~\ref{fig:rkw_tp}(d)], the ARPES can observe the major part of the pseudogap, in contrast to the hole-doped cases.
Around the node, the EDC peak position saturates at $\w\sim-0.05$eV for $Ln=$Eu compound, implying the pseudogap at $\omega \gtrsim -0.05$eV, while the pseudogap is not clear for $Ln=$Nd in EDC.
This is consistent with the observation in Fig.~\ref{fig:rkw_tp} that the pseudogap position in the nodal region is lowered for smaller $|t'/t|$ and that the Eu compound appears to have a relatively small $|t'/t|$. Namely, the main part of the pseudogap in the nodal region of the Nd compound is located in the positive energy side, while it becomes visible below $E_\text{F}$
 for the Eu compound.
This suggests a gap around the node, which can be interpreted as an indirect evidence of a gap opening above $E_\text{F}$ in hole-doped cases via the electron-hole transformation.
These observations are qualitatively consistent with the pole-zero structure in Fig.~\ref{fig:rkw_tp}(d), supporting the non-$d$-wave (fully-opened) pseudogap proposed in Sec.~\ref{ssec:nond}.
We propose to perform ARPES measurements of the electron doped compounds with a better resolution than Ref.~\onlinecite{iy09-2}, because the pseudogap appears to be suggested only by the jump in EDC and the detailed pseudogap structure in the momentum space is not very clear so far.

\begin{figure}[t]
\center{
\includegraphics[width=0.48\textwidth]{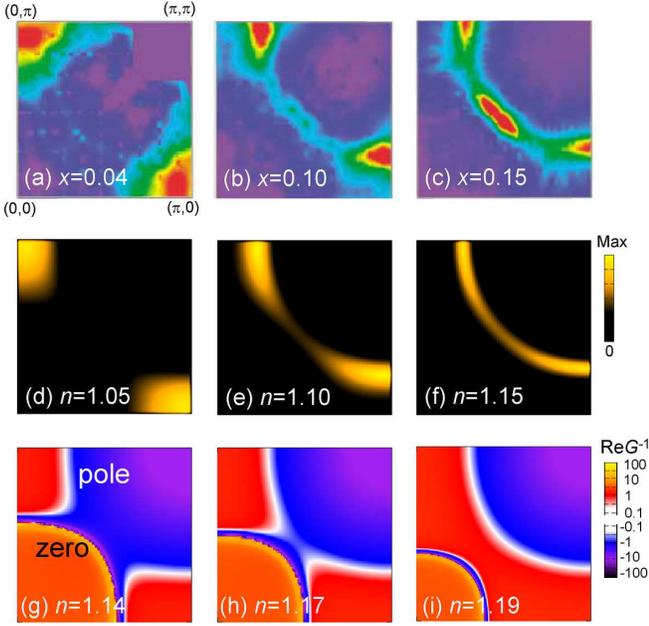}}
\caption{(color online). (a)-(c) ARPES spectra for Nd$_{2-x}$Ce$_x$CuO$_4$.
       Reproductions of Fig.~3 in Ref.~\onlinecite{ar02}.
       (d)-(f) $-\beta G(\Vec{k},\frac{\b}{2})$ calculated with 
       the CDMFT+CTQMC for $t'=-0.4t$, $U=12t$, and $T=0.033t$ at 
       $n=1.05$, $1.1$, and $1.15$, respectively.
       (g)-(i) Re$G^{-1}(\Vec{k},\w=0)$ calculated with the CDMFT+ED for
       $t'=-0.4t$, $U=12t$ at $n=1.14$, $1.17$, and $1.19$, respectively. 
       $\eta=0.0001t$ is used.
       }
\label{fig:edope}
\end{figure}

The ARPES \cite{ar02,iy09-2} has also revealed a characteristic evolution of the Fermi surface with doping.
Figures \ref{fig:edope}(a)-(c) are the reproductions of the ARPES results (Fig.~3 in Ref.~\onlinecite{ar02}) on Nd$_{2-x}$Ce$_x$CuO$_4$.
The ARPES study found that (i) at small doping ($x=0.04$) 
strong intensity emerges around $(\pi,0)$ and its symmetrically-related points [Fig.~\ref{fig:edope}(a)], (ii) as doping is increased, additional intensity develops around $(\frac{\pi}{2},\frac{\pi}{2})$ [Fig.~\ref{fig:edope}(b)], and (iii) at large doping the spectra merge, forming a single large surface around $(\pi,\pi)$ [Fig.~\ref{fig:edope}(c)].
In Ref.~\onlinecite{ar02} it is also seen that the band cutting the Fermi level at around $(\frac{\pi}{2},\frac{\pi}{2})$ evolves with doping from a high to low binding energy while that around $(\pi,0)$ emerges first around the Fermi level and extends to a high binding energy.
This suggests that the Fermi surface around $(\frac{\pi}{2},\frac{\pi}{2})$ is a hole pocket while the one around $(\pi,0)$ is an electron pocket.

The above features (i) and (iii) were qualitatively reproduced by the cluster perturbation theory \cite{st04} and by CDMFT with the $\S$ \cite{cc05} and $G$ \cite{kk06} periodizations on the 2D Hubbard model at $T=0$.
In Ref.~\onlinecite{st04} and \onlinecite{cc05} the suppression
of the intensity around the node at low dopings was understood by the presence of the hot spot, where Im$\S$ is large, in this region.

In the following we clarify how the above observations (i)-(iii) can be understood in terms of the underlying zero surface.
First, we show the CDMFT+CTQMC results at $T=0.05t$.
Figures \ref{fig:edope}(d)-(f) depict the integrated low-energy spectra, Eq.~(\ref{eq:ghb}), for $t'=-0.4t$, $U=12t$, at $5$, $10$, and $15$\% electron dopings, respectively. 
At $n=1.05$ a strong intensity emerges only around the antinodal points, in accord with (i). A large Fermi surface around $(\pi,\pi)$ at $n=1.15$ agrees with (iii).
As to (ii), our results do not show strong intensity around $(\frac{\pi}{2},\frac{\pi}{2})$; 
this will be attributed to the absence of the third-neighbor transfer integral $t''$ in the present calculation, as discussed below.

Second, Figs.~\ref{fig:edope}(g)-(i) show the CDMFT+ED results for $t'=-0.4t$ and $U=12t$ at $n=1.14$, $1.17$, and $1.19$, respectively.
We see that a zero surface exists at the Fermi level, surrounding $(0,0)$.
For low dopings electron-pocket Fermi surfaces reside around the antinodal points.
As the doping is increased, the pockets expand.
Then they merge around the nodal point and change the topology into two Fermi surfaces, one around $(\pi,\pi)$ and the other around $(0,0)$. 
Further doping makes the inner Fermi surface annihilate in pair with the zero surface, resulting in the Fermi liquid with the single large Fermi surface around $(\pi,\pi)$.\cite{footnote3}
Thus in electron doping there occur, at least, two phase transitions, the Lifshitz transition \cite{ko07} and a pole-zero annihilation transition, on the way from the Mott insulator to the Fermi liquid.

The presence of the zero surface accounts for the weak spectral intensity around the nodal direction in Figs.~\ref{fig:edope}(d) and (e).
Then how is the intensity around the node in Fig.~\ref{fig:edope}(b) and (c) 
explained?
To answer this we consider how the electronic structure, Fig.~\ref{fig:rkw_tp}(d), changes with $t''$, which is not included in our calculations based on the $2\times2$ cluster.
Although $|t''|$ is usually several times smaller than $|t'|$ and therefore negligible for discussing global electronic structures, it can be crucial to understand the nodal intensity.
This is because around $(\frac{\pi}{2},\frac{\pi}{2})$
the pole surface at $\w\simeq -0.15t$ in Fig.~\ref{fig:rkw_tp}(d) moves up with $-t'/t$, as seen in comparison with Fig.~\ref{fig:rkw_tp}(c), and is located just below $E_{\text{F}}$ for $-t'/t=0.4$ (not shown).
Since to the first approximation $t''(>0)$ elevates dispersions around $(\frac{\pi}{2},\frac{\pi}{2})$ while lowers around $(0,0)$, $(\pi,\pi)$, $(\pi,0)$, and $(0,\pi)$, according to
\begin{align}
\e(\Vec{k}) \rightarrow \e(\Vec{k})-2t''[\cos(2k_x)+\cos(2k_y)],
\end{align}
a hole pocket can emerge around $(\frac{\pi}{2},\frac{\pi}{2})$.
This hole pocket will show up just inside the zero surface in Figs.~\ref{fig:edope}(g) and (h).
Recent experimental observation \cite{hk09} of the Shubnikov-de Haas oscillation in Nd$_{2-x}$Ce$_x$CuO$_4$ also indicates a presence of the hole pocket around optimal doping.
Therefore, to confirm the above scenario with a larger cluster calculation incorporating $t''$ is an intriguing future problem.

We note some additional indications of the zero surface in experiments.
In Fig.~2(c)-(e) in Ref.~\onlinecite{ar02}, which plotted the ARPES EDCs for Nd$_{2-x}$Ce$_x$CuO$_4$ along the underlying Fermi surface, a clear jump of the peak positions can be seen between the nodal and antinodal directions.
This indicates the presence of the zero surface around the jump, in accord with the above picture [see Fig.~\ref{fig:rkw_tp}(d)].
In Fig.~1 in Ref.~\onlinecite{iy09-2}, which plotted the ARPES spectra at the Fermi level, a similar structure to Fig.~\ref{fig:edope}(c) was found while it clearly shows a finite intensity in some regions on the $(\pi,0)\text{-}(0,\pi)$ line.
This appears to be inconsistent with the zero surface on the $(\pi,0)\text{-}(0,\pi)$ line assumed in Ref.~\onlinecite{kr06} for hole-doped cases, but compatible with our zero surface around $(0,0)$ as seen in Figs.~\ref{fig:edope}(g) and (h).

\section{SUMMARY AND CONCLUSION}\label{sec:summary}

In summary, we have shown that various spectral anomalies
observed in the pseudogap states of hole- or electron-doped cuprates are naturally understood in terms of underlying pole-zero structure of electronic Green's function.
The pole-zero structure has been calculated for the paramagnetic metallic phase in the 2D Hubbard model with employing the CDMFT+ED at $T=0$ and the CDMFT+CTQMC at $T>0$.
We have confirmed that the cumulant periodization scheme \cite{sk06}
is suited for the parameter region of our interest, namely lightly doped Mott insulator in the intermediate to strong coupling region, 
since the cumulant is well localized within the cluster that we employed, 
as demonstrated in APPENDIX B. 
This corroborates the convergence and reliability of our momentum resolution in the studies on Fermi surface topology and the differentiation in the momentum space.
Furthermore, the cumulant periodization is useful 
since it enables to study the coexistence of pole and zero surfaces. 

The result calculated by CDMFT+ED shows a pseudogap in the lightly doped region around the Mott insulator. 
The pseudogap is characterized by a low-energy zero surface, which connects the bifurcated low-energy bands [Fig.~\ref{fig:fig2}(a)].
In hole-doped cases the lower band cuts the Fermi level while the upper one resides at the energy higher than and adjacent to the pseudogap above the Fermi level.
This directly leads to (i) the non-$d$-wave character of the fully opened pseudogap and (ii) the spectral asymmetry around the Fermi level.
Looking into the structure around $(\pi,\pi)$ we have also found that a large Re$\S$ around the zero surface causes (iii) the back-bending dispersion.
All of (i), (ii), and (iii) are consistent with experiments semiquantitatively.
The agreement imposes a strong constraint on theories for pseudogap:
Although (ii) and (iii) have been interpreted as evidences of the preformed pairing in the literatures, (i) is in sharp contrast to $d$-wave-gap scenarios including those by preformed pair.  
It is compatible neither with other precursory or real $d$-wave-type symmetry breakings such as $d$-density wave nor with commensurate antiferromagnetism. 
We have confirmed a full gap formation at the nodal point in a larger-cluster calculation in APPENDIX C. 
Moreover, the non-$d$-wave gap is supported by the agreement of our result on the pseudogap structure with that observed by ARPES for electron-doped cuprates.

The zero mechanism also enables a simple and unified understanding of 
various spectral anomalies:
In hole-doped cases the same zero surface which causes the back-bending behavior around the antinode induces the incoherence around $(\pi,\pi)$, Fermi pockets and Fermi arcs, while a pile of pole and zero surfaces at a higher binding energy results in the high-energy kink (waterfall) and the incoherence around $(0,0)$.
Moreover we have found a low-energy kink structure in the nodal dispersion.
On the other hand, in electron-doped cases the zero surface which constitutes the pseudogap crosses the Fermi level around $(0,0)$, making electron pockets around the antinodes.
All these features are consistent with experimental results. 
We would like to emphasize that the zero surface does not result from symmetry
breakings, but is a direct consequence of the strong correlation effect, i.e., proximity to the Mott insulator.

To elucidate the effect of zeros at finite temperatures, we have also implemented CDMFT+CTQMC calculations.
The results qualitatively agree with those of CDMFT+ED with 
a smearing factor $\eta$, which confirms that the thermal scatterings broaden the effect of zeros and indeed create the characteristic spectra similar to those observed in the cuprates.

To conclude, the origins of various anomalies in the electronic structure of the normal state in the high-$T_{\text{c}}$ cuprates are unified into 
the presence of the low-energy zero surface which persists against doping.
The zero surface is also expected to cause anomalous metallic behaviors in other physical quantities, such as the specific heat, electronic resistivity, and Hall coefficient.\cite{ts99}
Comparisons of these quantities with experiments remain for future studies.

Although the microscopic mechanism creating the zero surface has not been discussed in this paper, it has emerged clearly as the proximity to the Mott insulator.
The similarity in the structure of the zero surfaces between the pseudogap and the Mott gap in the undoped system [see Fig.~1(a) in Ref.~\onlinecite{sm09}] also implies the Mott origin of the pseudogap.
At the same time, however, the pseudogap 
does not appear as that directly or continuously connected to the zero surface which forms the Mott gap in the undoped state. 
This is because, for example in the hole-doped case, the zero surface creating the fully-opened pseudogap is bounded above by the existence of low-energy excitations far below the upper Hubbard band (UHB). 
The reason why the ingap states are separated by the pseudogap from the main quasiparticle band is left for an important future issue.  
In Ref.~\onlinecite{sm09} we proposed a scenario in which 
the emergence of such a zero surface in the low-energy scale and the resultant pseudogap are a remnant of the binding of doubly occupied site (doublon) and empty site (holon) drastically weakened by the screening of Coulomb repulsions by mobile carriers. This has something to do neither with any symmetry breaking nor with its precursor. Recently, it has been proposed that a gap arising from hybridization between the quasiparticle and a composite fermion excitation is responsible for the pseudogap and the related zero surface.~\cite{yamaji10}    
Possible supplementary role of strong antiferromagnetic fluctuations are left for future studies.
Further study on the mechanism of pseudogap will be described elsewhere.

\section*{ACKNOWLEDGMENT}
S. S. thanks Masafumi Udagawa, Giorgio Sangiovanni and Alessandro Toschi, and 
M. I. thanks Youhei Yamaji for useful discussions.
S. S. is also grateful to Karsten Held 
for the hospitality.
This work is supported by a Grant-in-Aid for Science Research on Priority Area ``Physics of New Quantum Phases in Superclean Materials" (Grant No.17071003) from MEXT, Japan. 
S. S. is supported by Research Fellowship of the Japan Society for the Promotion of Science for Young Scientists.
The calculations are partly performed at the Supercomputer Center, ISSP, University of Tokyo.

\section*{APPENDIX A: Breakdown of $\S$ periodization for the Mott insulator}

Here we note that a standard periodization technique, which substitutes $Q=\S$ to Eq.~(\ref{eq:periodize}), cannot reproduce the Mott gap for large $U$.
Because Im$G=0$ in the Mott gap, Im$\S$ must be 0 or $\infty$ in the whole Brillouin zone.
However, this situation does never occur in Eq.~(\ref{eq:periodize}) with $Q=\S$ unless $^\forall i,j \in \text{C}, {\rm Im}\S_{ij}^{\text{C}}=0$ or $^\exists (ij), \S_{ij}^{\text{C}}=\infty$ for all $\w$ inside the gap, and these conditions are never satisfied as far as both $t$ and $U$ are nonzero and finite.

\begin{figure}[t]
\center{
\includegraphics[width=0.48\textwidth]{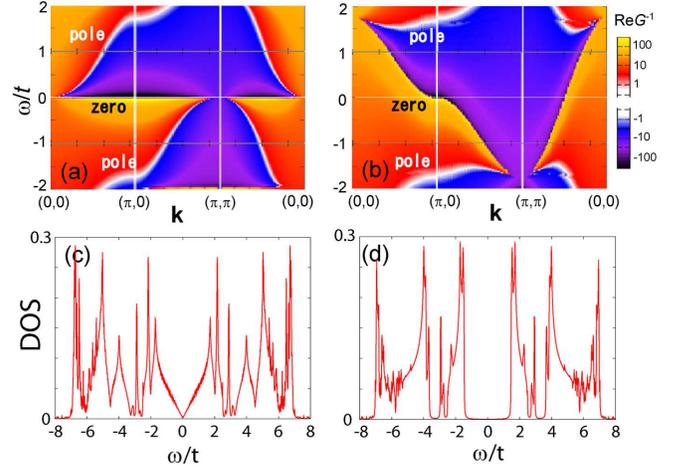}}
\caption{(color online). Re$G^{-1}(\Vec{k},\w)$ obtained with 
      (a) $\S$ and (b) $M$ periodization for a parameter region of 
      the Mott insulator ($t'=0$, $U=8t$, and $n=1$). 
      (c)(d) Density of states corresponding to (a) and (b), respectively.}
\label{fig:speriodize}
\end{figure}

As a matter of fact, we see the electronic structure shown in Fig.~\ref{fig:speriodize}(a) when we use the $\S$-periodization procedure for the same Mott insulator shown in Fig.~1(a) in Ref.~\onlinecite{sm09} [and reproduced in Fig.~\ref{fig:speriodize}(b) for comparison].
We see that zeros of $G$ exist only at the Fermi level without a dispersion, and that poles of $G$ extend to the Fermi level around $(0,0)$ from the positive $\w$ side and around $(\pi,\pi)$ from the negative $\w$ side, making the density of states half metallic.
This failure of the $\S$ periodization in the Mott insulator is ascribed to the fact that $\S$ is not localized within the $2\times 2$ cluster.\cite{sk06}
The nonlocality of $\S$ is a direct consequence of the presence of dispersive zeros of $G$, i.e., momentum-dependent divergence of $\S$.

As we have discussed so far, zeros of $G$ still persist in doped Mott insulators up to a critical doping level beyond which the Fermi liquid emerges.
Therefore $\S$ should be highly nonlocal also in the non-Fermi-liquid region and the $\S$ periodization again breaks down there.
On the contrary, the cumulant $M$ is well localized within the $2\times 2$ cluster in this region, as we mentioned in Sec.~\ref{sec:method}.
In APPENDIX B we give another numerical evidence for the locality of the cumulant in doped Mott insulators.

\section*{APPENDIX B: Locality of cumulant in doped Mott insulators}

Here we present a CDMFT+QMC result for an $N_{\text{c}}=4\times 4$ cluster in a parameter region of doped Mott insulators.
We use the Hirsch-Fye algorithm \cite{hf86} and calculate the cluster cumulant $M^{\text{C}}$ for $t'=0$, $U=8t$, $n=0.91$, and $T=0.1t$.

\begin{figure}[t]
\center{
\includegraphics[width=0.48\textwidth]{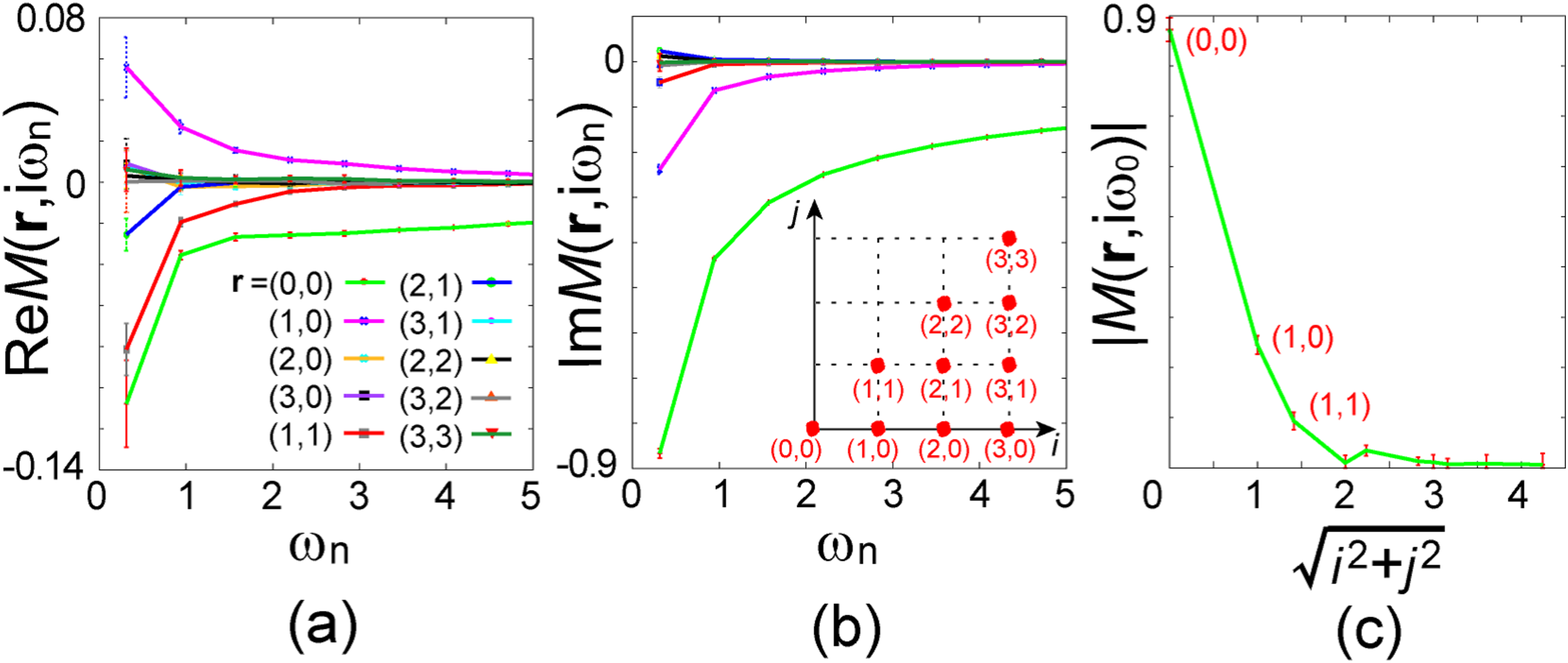}}
\caption{(color online). (a)(b) Matsubara-frequency dependence of the real and           imaginary parts of cumulants $M$ for various real-space vectors 
        $\Vec{r}=(i,j)$ in the $4\times 4$ cluster
        [shown in the inset of (b)], calculated with the
       CDMFT+QMC for $t'=0$, $U=8t$, $n=0.91$, and $T=0.1t$.
       (c) The cumulants at the lowest Matsubara frequency
       plotted against the Euclidean distance.
       }
\label{fig:cum}
\end{figure}

Figures \ref{fig:cum}(a) and (b) show $\w_n$ dependence of the real and imaginary parts of the cumulant $M(\Vec{r},\w_n)$, respectively, where $\Vec{r} = (i,j)$ is the real-space vector connecting the sites $i$ and $j$ in the cluster [the inset of Fig.~\ref{fig:cum}(b)]. 
We notice that $\Vec{r}=(0,0)$, (1,0), and (1,1) components are much larger than the other components at longer distances.
This can be more clearly seen in Fig.~\ref{fig:cum}(c), where the cumulant at the lowest Matsubara frequency, $\w_0$, is plotted against the Euclidean distance, $\sqrt{i^2+j^2}$.
We find that the cumulant is well localized within the $2\times 2$ cluster,
which justifies the $M$ periodization.

Although the cumulant at $T=0$ might be more extended than that at $T=0.1t$, we do not have a tractable way to examine it for a cluster larger than $2\times 2$.
It is worthwhile, however, to note that the finite-temperature results obtained by our CDMFT+CTQMC calculation are consistent with the CDMFT+ED results with finite $\eta$'s, as discussed in Sec.~\ref{ssec:arc}. This fact implies that the $M$ periodization with $2\times 2$ cluster still remains a good approximation even at $T=0$.

\section*{APPENDIX C: Nodal spectra obtained by 8-site CDMFT+CTQMC}
\begin{figure}[t]
\center{
\includegraphics[width=0.45\textwidth]{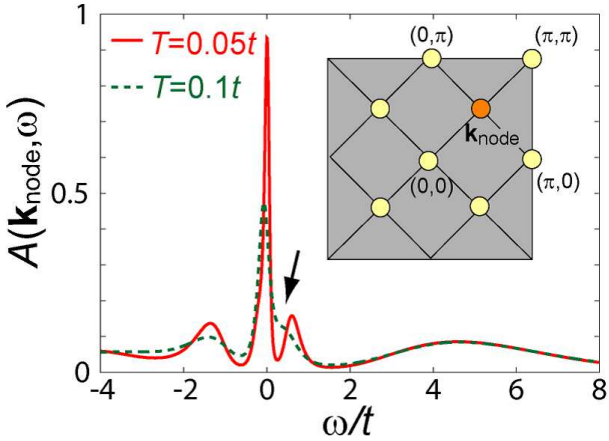}}
\caption{(color online). 
         Spectral functions at $\Vec{k}_\text{node}$,
         calculated with $N_\text{c}=8$ CDMFT+CTQMC for $t'=-0.2t$, $U=8t$, and
         $n=0.95$ at $T=0.05t$ (solid curve) and $0.1t$ (dashed curve),
         respectively. 
         The arrow indicates the gap discussed in the text.
         Inset: The momenta (circles) specifying the superlattice consisting of
         the 8-site clusters.}
\label{fig:8site}
\end{figure}
To demonstrate the fully-opened pseudogap in a larger cluster, we implement an 8-site cluster calculation with CDMFT+CTQMC.
We calculate the spectral function $A(\Vec{k},\w)$ at $\Vec{k}=\Vec{k}_\text{node}\equiv(\pi/2,\pi/2)$, where the $d$- and $s$-wave (fully-opened) pseudogaps are most distinguishable.
The spectral function is obtained from $G(\Vec{k},i\w_n)$ through the maximum-entropy method.

Figure \ref{fig:8site} shows $A(\Vec{k}_\text{node},\w)$ for $t'=-0.2t$, $U=8t$, and $n=0.95$.
The small but clear gap just above the Fermi level at $T=0.05t$ proves the fully-opened pseudogap.\cite{footnote4}
The gap vanishes at a higher temperature $T=0.1t$.
This is why the previous DCA study \cite{mj06} at a high temperature $T=0.12t$ did not observe the gap.
Note that, a very recent study with 8-site DCA \cite{lg10} at $T=0.05t$ also detected a wispy reduction of the spectral weight at $\Vec{k}_\text{node}$ at low doping though the signal was so weak that it was not explicitly analyzed in their paper.
The weakness of the signal would be due to the flat momentum dependence of the self-energy in each momentum patch in the DCA.
Although the suppression of the spectrum is more subtle at higher doping in their result, it is reasonable because the pseudogap temperature decreases with doping. 

The gap formation at the nodal point cannot be reproduced by the assumption of the $d$-wave pseudogap.
We have confirmed this by employing the assumed form of the Green function employed in Ref.~\onlinecite{kr06}.


\begin{references}
\item[\nonumber]
\bibitem{ts99}
T. Timusk and B. Statt, Rep. Prog. Phys. {\bf 62}, 61 (1999).
\bibitem{dh03}
A. Damascelli, Z. Hussain, and Z.-X. Shen, Rev. Mod. Phys. {\bf 75}, 473 (2003).
\bibitem{yj03}
Y. Yanase, T. Jujo, T. Nomura, H. Ikeda, T. Hotta, and K. Yamada, Phys. Rep. {\bf 387}, 1 (2003).
\bibitem{ek95}
V. J. Emery and S. A. Kivelson, Nature (London) {\bf 374}, 434 (1995).
\bibitem{ks90}
A. Kampf and J. R. Schrieffer, Phys. Rev. B {\bf 41}, 6399 (1990).
\bibitem{p97}
D. Pines, Z. Phys. B {\bf 103}, 129 (1997).
\bibitem{vt97}
Y. M. Vilk and A.-M. S. Tremblay, J. Phys. I France {\bf 7}, 1309 (1997).
\bibitem{kb03}
S. A. Kivelson, I. P. Bindloss, E. Fradkin, V. Oganesyan, J. M. Tranquada, A. Kapitulnik, and C. Howald, Rev. Mod. Phys. {\bf 75}, 1201 (2003).
\bibitem{v97}
C. M. Varma, Phys. Rev. B {\bf 55}, 14554 (1997).
\bibitem{cl01}
S. Chakravarty, R. B. Laughlin, D. K. Morr, and C. Nayak, Phys. Rev. B {\bf 63}, 094503 (2001).
\bibitem{rice05}
T. M. Rice, Prog. Theor. Phys. Suppl. {\bf 160}, 39 (2005).
\bibitem{sp03}
T. D. Stanescu and P. Phillips, Phys. Rev. Lett. {\bf 91}, 017002 (2003).
\bibitem{sk06}
T. D. Stanescu and G. Kotliar, Phys. Rev. B {\bf 74}, 125110 (2006);
T. D. Stanescu, M. Civelli, K. Haule, and G. Kotliar, Ann. Phys. (N. Y.) {\bf 321}, 1682 (2006).
\bibitem{p09}
P. Philips, T.-P. Choy, and R. G. Leigh, Rep. Prog. Phys. {\bf 72}, 036501 (2009).
\bibitem{sm09}
S. Sakai, Y. Motome, and M. Imada, Phys. Rev. Lett. {\bf 102}, 056404 (2009).

\bibitem{gk96}
A. Georges, G. Kotliar, W. Krauth, and M. J. Rozenberg, Rev. Mod. Phys. {\bf 68}, 13 (1996).
\bibitem{ks01}
G. Kotliar, S. Y. Savrasov, G. Palsson, and G. Biroli, Phys. Rev. Lett. {\bf 87}, 186401 (2001).
\bibitem{mj05}
T. Maier, M. Jarrell, T. Pruschke, and M. H. Hettler, Rev. Mod. Phys. {\bf 77}, 1027 (2005).
\bibitem{mp02}
Th. A. Maier, Th. Pruschke, and M. Jarrell, Phys. Rev. B {\bf 66}, 075102 (2002).
\bibitem{st04}
D. Se\'ne\'chal and A.-M. S. Tremblay, Phys. Rev. Lett. {\bf 92}, 126401 (2004).
\bibitem{cc05}
M. Civelli, M. Capone, S. S. Kancharla, O. Parcollet, and G. Kotliar, Phys. Rev. Lett. {\bf 95}, 106402 (2005).
\bibitem{kk06}
B. Kyung, S. S. Kancharla, D. Se\'ne\'chal, A.-M. S. Tremblay,M. Civelli, and G. Kotliar, Phys. Rev. B {\bf 73}, 165114 (2006).
\bibitem{mj06}
A. Macridin, M. Jarrell, T. Maier, P. R. C. Kent, and E. D'Azevedo, Phys. Rev. Lett {\bf 97}, 036401 (2006).
\bibitem{aa06}
M. Aichhorn, E. Arrigoni, M. Potthoff, and W. Hanke, Phys. Rev. B {\bf 74}, 024508 (2006).
\bibitem{hk07}
K. Haule and G. Kotliar, Phys. Rev. B {\bf 76}, 104509 (2007).
\bibitem{lt09}
A. Liebsch and N.-H. Tong, Phys. Rev. B {\bf 80}, 165126 (2009).
\bibitem{fc09}
M. Ferrero, P. S. Cornaglia, L. De Leo, O. Parcollet, G. Kotliar, and A. Georges, Phys. Rev. B {\bf 80}, 064501 (2009).

\bibitem{mj07}
A. Macridin, M. Jarrell, T. Maier, and D. J. Scalapino, Phys. Rev. Lett. {\bf 99}, 237001 (2007).
\bibitem{wg09}
P. Werner, E. Gull, O. Parcollet, and A. J. Millis, Phys. Rev. B {\bf 80}, 045120 (2009);
E. Gull, O. Parcollet, P. Werner, and A. J. Millis, {\it ibid} {\bf 80}, 245102 (2009).
\bibitem{sm09-2}
S. Sakai, Y. Motome, and M. Imada, Physica B {\bf 404}, 3183 (2009).
%%%%%%%% zero
\bibitem{d03}
I. Dzyaloshinskii, Phys. Rev. B {\bf 68}, 085113 (2003).
\bibitem{sp07}
T. D. Stanescu, P. Phillips, and T.-P. Choy, Phys. Rev. B {\bf 75}, 104503 (2007).
\bibitem{kr06}
R. M. Konik, T. M. Rice, and A. M. Tsvelik, Phys. Rev. Lett. {\bf 96}, 086407 (2006); 
K.-Y. Yang, T. M. Rice, and F.-C. Zhang, Phys. Rev. B {\bf 73}, 174501 (2006);
K.-Y. Yang, H.-B. Yang, P. D. Johnson, T. M. Rice, and F.-C. Zhang, Europhys. Lett. {\bf 86}, 37002 (2009); 
K. Le Hur and T. M. Rice, Ann. Phys. {\bf 324}, 1452 (2009).
\bibitem{ks06}
E. Z. Kuchinskii and M. V. Sadovskii, JETP {\bf 103}, 415 (2006). 
\bibitem{et02}
F. H. L. Essler and A. M. Tsvelik, Phys. Rev. B {\bf 65}, 115117 (2002); 
{\bf 71}, 195116 (2005).
\bibitem{bg06}
C. Berthod, T. Giamarchi, S. Biermann, and A. Georges, Phys. Rev. Lett. {\bf 97}, 136401 (2006).
\bibitem{hs90}
M. S. Hybertsen, E. B. Stechel, M. Schluter, and D. R. Jennison, Phys. Rev. B {\bf 41}, 11068 (1990).

%%%%%%%% CDMFT
%%%%%%% ED
\bibitem{l50}
C. Lanczos, J. Research NBS {\bf 45}, 255 (1950).
\bibitem{gb87}
E. R. Gagliano and C. A. Balseiro, Phys. Rev. Lett. {\bf 59}, 2999 (1987).
%%%%%%% CTQMC
\bibitem{rh99}
S. M. A. Rombouts, K. Heyde, and N. Jachowicz, Phys. Rev. Lett. {\bf 82}, 4155 (1999).
\bibitem{sa06}
S. Sakai, R. Arita, and H. Aoki, Physica B {\bf 378-380}, 288 (2006);
S. Sakai, R. Arita, K. Held, and H. Aoki, Phys. Rev. B {\bf 74}, 155102 (2006).\bibitem{gw08}
E. Gull, P. Werner, O. Parcollet, and M. Troyer, Eur. Phys. Lett. {\bf 82}, 57003 (2008).
\bibitem{rh98}
S. Rombouts, K. Heyde, and N. Jachowicz, Phys. Lett. A {\bf 242}, 271 (1998).
\bibitem{h83}
J. E. Hirsch, Phys. Rev. B {\bf 28}, 4059 (1983); {\bf 29}, 4159 (1984).
%%%%%%% non-d wave gap
\bibitem{bf98}
L. Balents, M. P. A. Fisher, and C. Nayak, Int. J. Mod. Phys. B {\bf 12}, 1033 (1998).
\bibitem{ft01}
M. Franz and Z. Tesanovic, Phys. Rev. Lett. {\bf 87}, 257003 (2001).
\bibitem{nd98}
M. R. Norman, H. Ding, M. Randeria, J. C. Campuzano, T. Yokoya, T. Takeuchi, T. Takahashi, T. Mochiku, K. Kadowaki, P. Guptasarma, and D. G. Hinks, Nature {\bf 392}, 157 (1998).
%%%%%%% e-h asym
\bibitem{hl04}
T. Hanaguri, C. Lupien. Y. Kohsaka, D.-H. Lee, M. Azuma, M. Takano, H.Takagi, and J. C. Davis, Nature {\bf 430}, 1001 (2004).
\bibitem{yr08}
H.-B. Yang, J. D. Rameau, P. D. Johnson, T. Valla, A. Tsvelik, and G. D. Gu, Nature {\bf 456}, 77 (2008).
\bibitem{bc57}
J. Bardeen, L. N. Cooper, and J. R. Schrieffer, Phys. Rev. {\bf 108}, 1175 (1957).
\bibitem{es75}
A. L. Efros and B. I. Shklovskii, J. Phys. C {\bf 8}, L49 (1975).
\bibitem{si09}
H. Shinaoka and M. Imada, Phys. Rev. Lett. {\bf 102}, 016404 (2009);
J. Phys. Soc. Jpn. {\bf 78}, 094708 (2009).
%%%%%%% back bending
\bibitem{kc08}
A. Kanigel, U. Chatterjee, M. Randeria, M. R. Norman, G. Koren, K. Kadowaki, and J. C. Campuzano, Phys. Rev. Lett. {\bf 101}, 137002 (2008).
\bibitem{hh10}
M. Hashimoto, R.-H. He, K. Tanaka, J.-P. Testaud, W. Meevasana, R. G. Moore, 
D. Lu, H. Yao, Y. Yoshida, H. Eisaki, T. P. Devereaux, Z. Hussain, and Z.-X. Shen, Nature Phys. NPHYS1632 (2010).
\bibitem{ph97}
R. Preuss, W. Hanke, C. Gro\"ber, and H. G. Evertz, Phys. Rev. Lett. {\bf 79}, 1122 (1997).
%%%%%%% edc
\bibitem{yz07}
T. Yoshida, X. J. Zhou, D. H. Lu, S. Komiya, Y. Ando, H. Eisaki, T. Kakeshita, S. Uchida, Z. Hussain, Z.-X. Shen, and A. Fujimori, J. Phys.: Condens. Matter {\bf 19}, 125209 (2007).
%%%%%%% waterfall
\bibitem{kb05}
A. A. Kordyuk, S. V. Borisenko, A. Koitzsch, J. Fink, M. Knupfer, and H. Berger, Phys. Rev. B. {\bf 71}, 214513 (2005).
\bibitem{gg07}
J. Graf, G.-H. Gweon, K. McElroy, S. Y. Zhou, C. Jozwiak, E. Rotenberg, A. Bill, T. Sasagawa, H. Eisaki, S. Uchida, H. Takagi, D.-H. Lee, and A. Lanzara, Phys. Rev. Lett. {\bf 98}, 067004 (2007).
\bibitem{xy07}
B. P. Xie, K. Yang, D. W. Shen, J. F. Zhao, H. W. Ou, J. Weil, S. Y. Gu, M. Arita, S. Qiao, H. Namatame, M. Taniguchi, N. Kaneko, H. Eisaki, K. D. Tsuei, C. M. Cheng, I. Vobornik, J. Fujii, G. Rossi, Z. Q. Yang, and D. L. Feng, Phys. Rev. Lett. {\bf 98}, 147001 (2007).
\bibitem{vk07}
T. Valla, T. E. Kidd, W.-G. Yin, G. D. Gu, P. D. Johnson, Z.-H. Pan, and A. V. Fedorov, Phys. Rev. Lett. {\bf 98}, 167003 (2007).
\bibitem{mz07}
W. Meevasana, X. J. Zhou, S. Sahrakorpi, W. S. Lee, W. L. Yang, K. Tanaka, N. Mannella, T. Yoshida, D. H. Lu, Y. L. Chen, R. H. He, Hsin Lin, S. Komiya, Y. Ando, F. Zhou, W. X. Ti, J. W. Xiong, Z. X. Zhao, T. Sasagawa, T. Kakeshita, K. Fujita, S. Uchida, H. Eisaki, A. Fujimori, Z. Hussain, R. S. Markiewicz, A. Bansil, N. Nagaosa, J. Zaanen, T. P. Devereaux, and Z.-X. Shen, Phys. Rev. B. {\bf 75}, 174506 (2007).
\bibitem{yt05}
T. Yoshida, K. Tanaka, H. Yagi, A. Ino, H. Eisaki, A. Fujimori, and Z.-X. Shen, Phys. Rev. Lett. {\bf 95}, 146404 (2005).
\bibitem{ec09}
R. Eguchi, A. Chainani, M. Taguchi, M. Matsunami, Y. Ishida, K. Horiba, Y. Senba, H. Ohashi, and S. Shin, Phys. Rev. B {\bf 79}, 115122 (2009).
\bibitem{bk07}
K. Byczuk, M. Kollar, K. Held, Y.-F. Yang, I. A. Nekrasov, Th. Pruschke, and D. Vollhardt, Nature Phys. {\bf 3}, 168 (2007).
%%%%%%% low-energy kink
\bibitem{bl00}
P. V. Bogdanov, A. Lanzara, S. A. Kellar, X. J. Zhou, E. D. Lu, W. J. Zheng, G. Gu, J.-I. Shimoyama, K. Kishio, H. Ikeda, R. Yoshizaki, Z. Hussain, and Z. X. Shen, Phys. Rev. Lett. {\bf 85}, 2581 (2000).
\bibitem{kr01}
A. Kaminski, M. Randeria, J. C. Campuzano, M. R. Norman, H. Fretwell, J. Mesot, T. Sato, T. Takahashi, and K. Kadowaki, Phys. Rev. Lett. {\bf 86}, 1070 (2001).\bibitem{sm03}
T. Sato, H. Matsui, T. Takahashi, H. Ding, H.-B. Yang, S.-C.Wang, T. Fujii, T. Watanabe, A. Matsuda, T. Terashima, and K. Kadowaki, Phys. Rev. Lett. {\bf 91}, 157003 (2003).
\bibitem{lb01}
A. Lanzara, P. V. Bogdanov, X. J. Zhou, S. A. Kellar, D. L. Feng, E. D. Lu, T. Yoshida, H. Eisaki, A. Fujimori, K. Kishio, J.-I. Shimoyama, T. Nodak, S. Uchidak, Z. Hussain, and Z.-X. Shen, Nature {\bf 412}, 510 (2001).
\bibitem{jv01}
P. D. Johnson, T. Valla, A. V. Fedorov, Z. Yusof, B. O. Wells, Q. Li, A. R. Moodenbaugh, G. D. Gu, N. Koshizuka, C. Kendziora, Sha Jian, and D. G. Hinks, Phys. Rev. Lett. {\bf 87}, 177007 (2001).
%%%%%%% hole pocket
\bibitem{footnote2}
Note that ARPES spectra are also usually integrated around the Fermi level and that a typical width of the energy window is $\sim 10\text{meV} \sim 0.025t$, which is similar to the $T$ window used in Eq.~(\ref{eq:ghb}).
\bibitem{gw08-2}
E. Gull, P. Werner, X. Wang, M. Troyer, and A. J. Millis, Europhys. Lett. {\bf 84}, 37009 (2008).
\bibitem{ml09}
J. Meng, G. Liu, W. Zhang, L. Zhao, H. Liu, X. Jia, D. Mu, S. Liu, X. Dong, J. Zhang, W. Lu, G. Wang, Y. Zhou, Y. Zhu, X. Wang, Z. Xu, C. Chen, and X. J. Zhou, Nature {\bf 462}, 335 (2009).
%%%%%%% electron dope
\bibitem{iy09}
M. Ikeda, T. Yoshida, A. Fujimori, M. Kubota, K. Ono, Y. Kaga, T. Sasagawa, and H. Takagi, Phys. Rev. B. {\bf 80}, 184506 (2009).
\bibitem{footnote5}
The spectral feature that the band bottom is at a higher energy around $(\pi,0)$ than around $(0,0)$, as well as a waterfall-like structure at $\w>0$, can also be seen in Fig.~5(b) in Ref.~\onlinecite{kk06}.
\bibitem{ot01}
Y. Onose, Y. Taguchi, K. Ishizaka, and Y. Tokura, Phys. Rev. Lett. {\bf 87}, 217001 (2001).
\bibitem{kr03}
H. Kusunose and T. M. Rice, Phys. Rev. Lett. {\bf 91}, 186407 (2003).
\bibitem{kh04}
B. Kyung, V. Hankevych, A.-M. Dare\', and A.-M. S. Tremblay, Phys. Rev. Lett. {\bf 93}, 147004 (2004).
\bibitem{iy09-2}
M. Ikeda, T. Yoshida, A. Fujimori, M. Kubota, K. Ono, H. Das, T. Saha-Dasgupta, K. Unozawa, Y. Kaga, T. Sasagawa, and H. Takagi, Phys. Rev. B {\bf 80}, 014510 (2009).
\bibitem{ar02}
N. P. Armitage, F. Ronning, D. H. Lu, C. Kim, A. Damascelli, K. M. Shen, D. L. Feng, H. Eisaki, Z.-X. Shen, P. K. Mang, N. Kaneko, M. Greven, Y. Onose, Y. Taguchi, and Y. Tokura, Phys. Rev. Lett. {\bf 88}, 257001 (2002).
\bibitem{footnote3}
The doping evolution of the Fermi surface and zero surface in hole-doped cases was discussed in Ref.~\onlinecite{sm09}.
\bibitem{ko07}
M. M. Korshunov and S. G. Ovchinnikov, Eur. Phys. J. B {\bf 57}, 271 (2007);
M. M. Korshunov, E. V. Zakharova, I. A. Nekrasov, Z. V. Pchelkina, and S. G. Ovchinnikov, J. Phys.: Condens. Matter {\bf 22}, 015701 (2010).
\bibitem{hk09}
T. Helm, M. V. Kartsovnik, M. Bartkowiak, N. Bittner, M. Lambacher, A. Erb, J. Wosnitza, and R. Gross, Phys. Rev. Lett. {\bf 103}, 157002 (2009).
%%%%%%% summary
\bibitem{lg10}
N. Lin, E. Gull, and A. J. Millis, arXiv:1004.2999.
\bibitem{yamaji10}
Y. Yamaji and M. Imada, unpublished.
%%%%%%% appendix
\bibitem{hf86}
J. E. Hirsch and R. M. Fye, Phys. Rev. Lett. {\bf 56}, 2521 (1986).
\bibitem{footnote4}
The density of states also shows a gap at the corresponding energy.
\end{references}
\end{document}